\documentstyle[12pt,epsf]{article}

\newlength{\pubnumber} 

\catcode`\@=11
\@addtoreset{equation}{section}

\def\section{\@startsection{section}{1}{\z@}{3.5ex plus 1ex minus .2ex}
 {2.3ex plus .2ex}{\large\bf}}
\def\subsection{\@startsection{subsection}{2}{\z@}{2.3ex plus .2ex}
 {2.3ex plus .2ex}{\bf}}

\begin{document}

\begin{titlepage}
\samepage{
\setcounter{page}{1}
\rightline{ACT-9/99, CTP-TAMU-37/99}
\rightline{TPI-MINN-99/46, UMN-TH-1821-99}
\rightline{\tt hep-ph/9910230}
\rightline{October 1999}
\vskip .4truecm
\begin{center}
 {\Large \bf A Minimal Superstring Standard Model II:\\
   A Phenomenological Study\\ }
\vskip .4truecm
\vfill {\large
        G.B. Cleaver,$^{1,2}$\footnote{gcleaver@rainbow.physics.tamu.edu}
        A.E. Faraggi,$^{3}$\footnote{faraggi@mnhepw.hep.umn.edu}
        D.V. Nanopoulos,$^{1,2,4}$\footnote{dimitri@soda.physics.tamu.edu}
        and}

       {\large 
        J.W. Walker,$^{1}$\footnote{jwalker@rainbow.physics.tamu.edu}}
\\
\vspace{.12in}
{\it $^{1}$ Center for Theoretical Physics,
            Dept.\  of Physics, Texas A\&M University,\\
            College Station, TX 77843, USA\\}
\vspace{.06in}
{\it $^{2}$ Astro Particle Physics Group,
            Houston Advanced Research Center (HARC),\\
            The Mitchell Campus,
            Woodlands, TX 77381, USA\\}
\vspace{.06in}
{\it$^{3}$  Department of Physics, University of Minnesota, 
            Minneapolis, MN 55455, USA\\}
\vspace{.025in}
{\it$^{4}$  Academy of Athens, Chair of Theoretical Physics, 
            Division of Natural Sciences,\\
            28 Panepistimiou Avenue, Athens 10679, Greece\\}
\vspace{.025in}
\end{center}
\vfill
\begin{abstract}
Recently, we demonstrated the existence of heterotic--string solutions
in which the observable sector effective field theory just below 
the string scale reduces to that of the MSSM, 
with the standard observable gauge group being just
$SU(3)_C\times SU(2)_L\times U(1)_Y$ and the
$SU(3)_C\times SU(2)_L\times U(1)_Y$--charged 
spectrum of the observable sector consisting solely of the MSSM spectrum.
Associated with this model is a set of distinct flat directions of 
vacuum expectation values (VEVs) of non--Abelian singlet fields
that all produce solely the MSSM spectrum. 
In this paper, we study the effective superpotential
induced by these choices of flat directions. 
We investigate whether sufficient degrees of freedom exist in 
these singlet flat directions to satisfy various phenomenological 
constraints imposed by the observed Standard Model data. 
For each flat direction, the effective superpotential is given to
sixth order.  
The variations in the singlet and hidden sector low energy spectrums
are analyzed.  
We then determine the mass matrices (to all finite orders) for the
three generations of MSSM quarks and leptons.
Possible Higgs $\mu$--terms are investigated. 
We conclude by considering generalizations of our  
flat directions involving VEVs of non--Abelian fields.
\end{abstract}
\smallskip}
\end{titlepage}

\setcounter{footnote}{0}

\def\at{ }
\def\beq{\begin{equation}}
\def\eeq{\end{equation}}
\def\beqn{\begin{eqnarray}}
\def\eeqn{\end{eqnarray}}
\def\no{\noindent }
\def\nolabel{\nonumber }

\def\NA{non--Abelian }

\def\gsim{{\buildrel >\over \sim}}
\def\lsim{{\buildrel <\over \sim}}

\def\ie{i.e., }
\def\eg{{\it e.g.}}
\def\eq#1{eq.\ (\ref{#1})}

\def\lt{<}

\def\slash#1{#1\hskip-6pt/\hskip6pt}
\def\slk{\slash{k}}

\def\dag{\dagger}
\def\qandq{\quad {\rm and} \quad} 
\def\qand{\quad {\rm and} } 
\def\andq{ {\rm and} \quad } 
\def\qwithq{\quad {\rm with} \quad} 
\def\qwith{ \quad {\rm with} } 
\def\withq{ {\rm with} \quad} 

\def\fhalf{\frac{1}{2}}
\def\fsqrt{\frac{1}{\sqrt{2}}}
\def\half{{\textstyle{1\over 2}}}
\def\third{{\textstyle {1\over3}}}
\def\quarter{{\textstyle {1\over4}}}
\def\sixth{{\textstyle {1\over6}}}
\def\m{{\tt -}}
\def\ps{{\tt +}}
\def\pps{\phantom{+}}

\def\zz{$Z_2\times Z_2$ }

\def\Tr{{\rm Tr}\, }
\def\tr{{\rm tr}\, }

\def\MP{M_{P}}
\def\GeV{\,{\rm GeV}}
\def\TeV{\,{\rm TeV}}

\def\lam#1{\lambda_{#1}}
\def\non{\nonumber}
\def\smgg{ $SU(3)_C\times SU(2)_L\times U(1)_Y$ }
\def\smggb{ $SU(3)_C\times SU(2)_L\times U(1)_Y$}
\def\SM{Standard--Model }
\def\SUSY{supersymmetry }
\def\SSSM{supersymmetric standard model}
\def\MSSM{minimal supersymmetric standard model}
\def\MSSSM{MS$_{str}$SM }
\def\MSSSMc{MS$_{str}$SM, }
\def\obs{{\rm observable}}
\def\sig{{\rm singlets}}
\def\hid{{\rm hidden}}
\def\MS{M_{Pl}}
\def\Ms{$M_{str}$}

\def\vev#1{< #1 >}
\def\mvev#1{|< #1 >|^2}

\def\UA{U(1)_{\rm A}}
\def\QA{Q^{(\rm A)}}
\def\mssm{SU(3)_C\times SU(2)_L\times U(1)_Y} 

\def\KM{Ka\v c--Moody }

\def\y{\,{\rm y}}
\def\l{\langle}
\def\r{\rangle}
\def\o#1{\frac{1}{#1}}

\def\zi{z_{\infty}}
\def\zinf{z_{\infty}}

\def\hb#1{\bar{h}_{#1}}
\def\bh{\bar{h}}
\def\bhp{\bar{h^{'}}}
\def\hbo{\bar{h}}
\def\Htw{{\tilde H}}
\def\chibar{{\overline{\chi}}}
\def\qbar{{\overline{q}}}
\def\ibar{{\overline{\imath}}}
\def\jbar{{\overline{\jmath}}}
\def\Hbar{{\overline{H}}}
\def\Qbar{{\overline{Q}}}
\def\abar{{\overline{a}}}
\def\alphabar{{\overline{\alpha}}}
\def\betabar{{\overline{\beta}}}
\def\tautwo{{ \tau_2 }}
\def\thetatwo{{ \vartheta_2 }}
\def\thetathree{{ \vartheta_3 }}
\def\thetafour{{ \vartheta_4 }}
\def\ttwo{{\vartheta_2}}
\def\tthree{{\vartheta_3}}
\def\tfour{{\vartheta_4}}
\def\ti{{\vartheta_i}}
\def\tj{{\vartheta_j}}
\def\tk{{\vartheta_k}}
\def\calF{{\cal F}}
\def\smallmatrix#1#2#3#4{{ {{#1}~{#2}\choose{#3}~{#4}} }}
\def\ab{{\alpha\beta}}
\def\Minv{{ (M^{-1}_\ab)_{ij} }}
\def\ii{{(i)}}
\def\V{{\bf V}}
\def\N{{\bf N}}

\def\b{{\bf b}}
\def\S{{\bf S}}
\def\X{{\bf X}}
\def\I{{\bf I}}
\def\bone{{\mathbf 1}}
\def\bo{{\mathbf 0}}
\def\bs{{\mathbf S}}
\def\mS{{\mathbf S}}
\def\bS{{\mathbf S}}
\def\bb{{\mathbf b}}
\def\mb{{\mathbf b}}
\def\mX{{\mathbf X}}
\def\mI{{\mathbf I}}
\def\bI{{\mathbf I}}
\def\balpha{{\mathbf \alpha}}
\def\bbeta{{\mathbf \beta}}
\def\bgamma{{\mathbf \gamma}}
\def\bxi{{\mathbf \xi}}
\def\malpha{{\mathbf \alpha}}
\def\mbeta{{\mathbf \beta}}
\def\mgamma{{\mathbf \gamma}}
\def\mxi{{\mathbf \xi}}
\def\bphi{\overline{\Phi}}

\def\eps{\epsilon}

\def\t#1#2{{ \Theta\left\lbrack \matrix{ {#1}\cr {#2}\cr }\right\rbrack }}
\def\C#1#2{{ C\left\lbrack \matrix{ {#1}\cr {#2}\cr }\right\rbrack }}
\def\tp#1#2{{ \Theta'\left\lbrack \matrix{ {#1}\cr {#2}\cr }\right\rbrack }}
\def\tpp#1#2{{ \Theta''\left\lbrack \matrix{ {#1}\cr {#2}\cr }\right\rbrack }}
\def\l{\langle}
\def\r{\rangle}

\def\op#1{$\Phi_{#1}$}
\def\opp#1{$\Phi^{'}_{#1}$}
\def\opb#1{$\overline{\Phi}_{#1}$}
\def\opbp#1{${\overline{{\Phi}^{'}}_{#1}}$}
\def\oppb#1{${\overline{{\Phi}^{'}}_{#1}}$}
\def\oppx#1{$\Phi^{(')}_{#1}$}
\def\opbpx#1{${\overline{{\Phi}^{(')}}_{#1}}$}

\def\oh#1{$h_{#1}$}
\def\ohb#1{${\bar{h}}_{#1}$}
\def\ohp#1{$h^{'}_{#1}$}

\def\oQ#1{$Q_{#1}$}
\def\odc#1{$d^{c}_{#1}$}
\def\ouc#1{$u^{c}_{#1}$}

\def\oL#1{$L_{#1}$}
\def\oec#1{$e^{c}_{#1}$}
\def\oNc#1{$N^{c}_{#1}$}

\def\oH#1{$H_{#1}$}
\def\oV#1{$V_{#1}$}
\def\oHs#1{$H^{s}_{#1}$}
\def\oVs#1{$V^{s}_{#1}$}

\def\p#1{{\Phi_{#1}}}
\def\pp#1{{\Phi^{'}_{#1}}}
\def\pb#1{{{\overline{\Phi}}_{#1}}}
\def\pbp#1{{\overline{{\Phi}^{'}}_{#1}}}
\def\ppb#1{{\overline{{\Phi}^{'}}_{#1}}}
\def\ppx#1{{\Phi^{(')}_{#1}}}
\def\pbpx#1{{\overline{\Phi^{(')}}_{#1}}}

\def\h#1{h_{#1}}
\def\hb#1{{\bar{h}}_{#1}}
\def\hp#1{h^{'}_{#1}}

\def\Q#1{Q_{#1}}
\def\dc#1{d^{c}_{#1}}
\def\uc#1{u^{c}_{#1}}

\def\L#1{L_{#1}}
\def\ec#1{e^{c}_{#1}}
\def\Nc#1{N^{c}_{#1}}

\def\H#1{H_{#1}}
\def\V#1{V_{#1}}
\def\Hs#1{{H^{s}_{#1}}}
\def\Vs#1{{V^{s}_{#1}}}

\def\fdtv{FD2V }
\def\fdtp{FD2$^{'}$ }
\def\fdtpv{FD2$^{'}$v }

\def\FD2pv{FD2$^{'}$V }
\def\FD2p{FD2$^{'}$ }

\def\simlt{\stackrel{<}{{}_\sim}}
\def\simgt{\stackrel{>}{{}_\sim}}

\def\inbar{\,\vrule height1.5ex width.4pt depth0pt}

\def\IC{\relax\hbox{$\inbar\kern-.3em{\rm C}$}}
\def\IQ{\relax\hbox{$\inbar\kern-.3em{\rm Q}$}}
\def\IR{\relax{\rm I\kern-.18em R}}
 \font\cmss=cmss10 \font\cmsss=cmss10 at 7pt
\def\IZ{\relax\ifmmode\mathchoice
 {\hbox{\cmss Z\kern-.4em Z}}{\hbox{\cmss Z\kern-.4em Z}}
 {\lower.9pt\hbox{\cmsss Z\kern-.4em Z}}
 {\lower1.2pt\hbox{\cmsss Z\kern-.4em Z}}\else{\cmss Z\kern-.4em Z}\fi}

\def\AEF{A.E. Faraggi}
\def\AP#1#2#3{{\it Ann.\ Phys.}\/ {\bf#1} (19#2) #3}
\def\NPB#1#2#3{{\it Nucl.\ Phys.}\/ {\bf B#1} (19#2) #3}
\def\NPBPS#1#2#3{{\it Nucl.\ Phys.}\/ {{\bf B} (Proc. Suppl.) {\bf #1}} (19#2) 
 #3}
\def\PLB#1#2#3{{\it Phys.\ Lett.}\/ {\bf B#1} (19#2) #3}
\def\PRD#1#2#3{{\it Phys.\ Rev.}\/ {\bf D#1} (19#2) #3}
\def\PRL#1#2#3{{\it Phys.\ Rev.\ Lett.}\/ {\bf #1} (19#2) #3}
\def\PRT#1#2#3{{\it Phys.\ Rep.}\/ {\bf#1} (19#2) #3}
\def\PTP#1#2#3{{\it Prog.\ Theo.\ Phys.}\/ {\bf#1} (19#2) #3}
\def\MODA#1#2#3{{\it Mod.\ Phys.\ Lett.}\/ {\bf A#1} (19#2) #3}
\def\IJMP#1#2#3{{\it Int.\ J.\ Mod.\ Phys.}\/ {\bf A#1} (19#2) #3}
\def\nuvc#1#2#3{{\it Nuovo Cimento}\/ {\bf #1A} (#2) #3}
\def\RPP#1#2#3{{\it Rept.\ Prog.\ Phys.}\/ {\bf #1} (19#2) #3}
\def\etal{{\it et al\/}}

\def\hpx{h^{'}}
\def\hbp{\bar{h}^{'}}
\def\hpb{\bar{h}^{'}}

\hyphenation{su-per-sym-met-ric non-su-per-sym-met-ric}
\hyphenation{space-time-super-sym-met-ric}
\hyphenation{mod-u-lar mod-u-lar--in-var-i-ant}


\section{Minimal Superstring Standard Models} 

Recently \cite{cfn1,cfn2} we demonstrated that it is indeed possible 
for a string model \cite{fny,fc} to have exactly the 
minimal supersymmetric standard model (MSSM) fields
as the \smggb--charged matter content of its low energy effective 
field theory. 
We propose that string models with this property
be classified as ``Minimal Superstring Standard Models'' (\MSSSM) \cite{cfn2}.
In our \MSSSMc,  decoupling of all MSSM--charged exotics from
the low energy effective field theory 
was accomplished by sets of vacuum expectation values (VEVs) 
that eliminate the anomalous $U(1)_A$ endemic to  
several classes of string models (in particular, those of 
bosonic lattice, orbifold, or free fermionic construction).  
Besides restoring spacetime supersymmetry through cancellation of
the Fayet--Iliopoulos (FI) $D$--term, these sets of VEVs also  
give FI--scale ($\approx 4$ to $7\times 10^{16}$ GEV) mass to the MSSM
exotics. 
 
If the underlying ``initial'' state of the universe truly was 
an anomalous $U(1)_A$ string model, then  
determination of the specific flat VEV direction chosen to cancel 
the FI $D$--term was a result of non--perturbative dynamics.
That is, the physically preferred flat direction cannot be 
identified perturbatively. 
However, through perturbative means we can locate 
and classify the 
possible flat directions most consistent with 
observed data. Classification 
of  non--Abelian (NA) {\it singlet} flat directions
that produce the MSSM gauge group and matter fields, while
simultaneously decoupling all MSSM exotics,
was performed in \cite{cfn2}. 
Based on our ``stringent'' $F$--flatness constraints,
we found three directions flat to all order, one direction
flat to $12^{\rm th}$ order, and around 100 remaining directions
only flat to seventh order or less. 

The existence of free fermionic models with solely the MSSM spectrum
below the string scale reinforces the motivation to improve
our understanding of this particular class of string models.
Both from the point of view of understanding the non--perturbative
dynamics, as well as improving the techniques that are needed
in order to confront the perturbative string models with
the low energy experimental data. 
In this paper we perform studies of the phenomenological features of these 
first four flat directions
of the ``FNY'' model of \cite{fny,fc}.
We explore the phenomenology of our \MSSSM
flat directions and investigate
which (if any) of our four singlet directions appear most consistent with 
observed phenomenological criteria. 

We remark that phenomenological studies, similar to
the one performed in this paper, were done in the past
for other three generation free fermionic models. 
The new features in this paper are as follows.
First, the FNY model is the first known example
of a semi--realistic string model, which produces
solely the MSSM--charged spectrum just below the string scale.
Thus, for the first time such a phenomenological
analysis is carried out in a Minimal Superstring Standard Model.
Second, and more importantly, in the phenomenological
analysis performed in this paper, we implement the
systematic techniques for the analysis of F and D
flat directions that were developed over the last
few years \cite{dfset,cceel2,gcmh,cfn2}. Relative to the more
primitive studies performed in the past, our study here 
has the advantage that it incorporates in much of the analysis
the non--renormalizable terms to all finite orders. 
In the analysis we
systematically decouple from the effective low energy
field theory the fields that become superheavy and their
superpotential couplings.
Furthermore, the solutions that we study in this paper
are flat to all orders. We emphasize that such an 
exhaustive analysis is performed for the first time in a 
semi--realistic string model.
Thus, our paper further advances the methodology needed to
confront potentially viable superstring models with 
experimental data.

Our paper is organized as follows.
In Section 2 we present a brief review of 
$Z_2\times Z_2$ free fermionic models, 
the class from which the ``FNY'' model originates.
In Section 3  we review $D$-- and $F$--flatness constraints.
Our study and discussion of the phenomenology of our four flat directions 
appears in Section 4.
For each of the flat directions, we analyze the three generation mass
matrices for up, down, electron, and neutrino states, and study
the effective Higgs $\mu$ terms. 
We also investigate coupling constant strength
for high order superpotential terms. We conclude Section 4 with study of 
additional $F$--flatness constraints imposed when both fields in a vector--like
pair (with opposite $U(1)$ charges) acquire VEVs.
Lastly, in Section 5 we include some general discussion and
overview of our singlet flat directions. We  
briefly consider generalizations of them
in which NA fields are also allowed to take on VEVs.

\section{$Z_2\times Z_2$ free fermionic models}

Constructing minimal superstring models, {\it i.e.}
models with solely the MSSM spectrum below the string 
scale, is clearly the coveted goal of superstring phenomenology. 
However, it should be emphasized that the success of
the FNY model in achieving this goal should not be viewed
as indicating that the model of ref.\  \cite{fny} is the correct
string vacuum. In this respect it is important to
understand that the FNY model belongs to a large class
of three generation free fermionic models, which possess
an underlying $Z_2\times Z_2$ orbifold structure.
Many of the issues pertaining to the phenomenology
of the Standard Model and supersymmetric unification
have been addressed in the past in the framework
of the quasi--realistic free fermionic models. An important
property of these models is the fact that they produce 
three generation models with the standard $SO(10)$ embedding of the 
Standard Model spectrum. No other orbifold
heterotic--string compactification has yielded a similar
structure. The FNY model should be viewed as a prototype
example of a semi--realistic free fermionic model.
The success of the FNY model in producing solely 
the MSSM spectrum below the string scale should
then be regarded as providing further evidence for the
assertion that the true string vacuum is connected
to the $Z_2\times Z_2$ orbifold in the vicinity of the 
free fermionic point in the Narain moduli space.

For completeness we recall the basic structure of
the free fermionic superstring models. The purpose 
is to highlight the fact that the free fermionic
models correspond to a large set of viable three generation
models, which differ in their detailed phenomenological
characteristics. In this respect, the FNY model
should be regarded as a representative example. 

A model in the free fermionic
formulation \cite{fff} is defined by a set of  boundary
condition basis vectors, and one--loop GSO phases, which are
constrained by the string consistency requirements,
and which completely determine the vacuum structure of the models.
The physical spectrum is obtained by applying the generalized 
GSO projections.

The first five basis vectors of the $Z_2\times Z_2$ free fermionic
models consist of the NAHE set \cite{nahe,eu,foc}.
The gauge group after the NAHE set is
$SO(10)\times E_8\times SO(6)^3$ with $N=1$ space--time supersymmetry, 
and 48 spinorial $\mathbf 16$'s of $SO(10)$, sixteen from each sector $\mb_1$,
$\mb_2$ and $\mb_3$. The three sectors $\mb_1$, $\mb_2$ and $\mb_3$ are
the three twisted sectors of the corresponding $Z_2\times Z_2$
orbifold compactification. The $Z_2\times Z_2$ orbifold is special
precisely because of the existence of three twisted sectors,
with a permutation symmetry with respect to the horizontal $SO(6)^3$
symmetries. The NAHE set is depicted in the table below which
highlights its cyclic permutation symmetry.
The NAHE set is common to a large class of three generation
free fermionic models. The construction proceeds by adding to the
NAHE set three additional boundary condition basis vectors
which break $SO(10)$ to one of its subgroups, $SU(5)\times U(1)$,
$SO(6)\times SO(4)$ or $SU(3)\times SU(2)\times U(1)^2$,
and at the same time reduces the number of generations to
three, one from each of the sectors $\mb_1$, $\mb_2$ and $\mb_3$.
The various three generation models differ in their
detailed phenomenological properties. These detailed properties
depend on the specific assignment of boundary condition
basis vector for the internal world--sheet fermions
$\{y,\omega\vert{\bar y},{\bar\omega}^{1,\cdots,6}\}$.
However, many of
the characteristics of the three generation models
can be traced back to the underlying
NAHE set structure. One such important property to note
is the fact that, as the three generations are obtained
from the three twisted sectors $\mb_1$, $\mb_2$ and $\mb_3$,
they automatically possess the Standard $SO(10)$ embedding.
Consequently, the weak hypercharge, which arises as
the usual combination $U(1)_Y= U(1)_{T_{3_R}} + \frac{1}{2} U(1)_{B-L}$,
has the standard $SO(10)$ embedding. To date, of the three generation
heterotic--orbifold models that have been constructed, only the
free fermionic models have yielded such a structure.

\vspace{0.15cm}

~~~~~{{\bf THE NAHE SET}}
{\large
{
\beqn
 &&\begin{tabular}{c|c|ccc|c|ccc|c}
 ~ & $\psi^\mu$ & ${\chi^{12}}$ & ${\chi^{34}}$ & ${\chi^{56}}$ &
        $\bar{\psi}^{1,...,5} $ &
        {$\bar{\eta}^1$}&
        {$\bar{\eta}^2$}&
        {$\bar{\eta}^3$}&
        $\bar{\phi}^{1,...,8} $ \\
\hline
\hline
      {\bf 1} &  1 & 1&1&1 & 1,...,1 & 1 & 1 & 1 & 1,...,1 \\
        $\mS$ &  1 & {1}&{1}&{1} & 0,...,0 & 0 & 0 & 0 & 0,...,0 \\
\hline
  {${\mb}_1$} &  1 & {1}&0&0 & 1,...,1 & {1} & 0 & 0 & 0,...,0 \\
  {${\mb}_2$} &  1 & 0&{1}&0 & 1,...,1 & 0 & {1} & 0 & 0,...,0 \\
  {${\mb}_3$} &  1 & 0&0&{1} & 1,...,1 & 0 & 0 & {1} & 0,...,0 \\
\end{tabular}
   \nonumber\\
   ~  &&  ~ \nonumber\\
     &&\begin{tabular}{c|cc|cc|cc}
 ~&     {$y^{3,...,6}$}  &
        {${\bar y}^{3,...,6}$}  &
        {$y^{1,2},\omega^{5,6}$}  &
        {${\bar y}^{1,2},\bar{\omega}^{5,6}$}  &
        {$\omega^{1,...,4}$}  &
        {$\bar{\omega}^{1,...,4}$}   \\
\hline
\hline
    {\bf 1} & 1,...,1 & 1,...,1 & 1,...,1 & 1,...,1 & 1,...,1 & 1,...,1 \\
   $\mS$    & 0,...,0 & 0,...,0 & 0,...,0 & 0,...,0 & 0,...,0 & 0,...,0 \\
\hline
{${\mb}_1$} & {1,...,1} & {1,...,1} & 0,...,0 & 0,...,0 & 
                                          0,...,0 & 0,...,0 \\
{${\mb}_2$} & 0,...,0 & 0,...,0 & {1,...,1} & {1,...,1} & 
                                          0,...,0 & 0,...,0 \\
{${\mb}_3$} & 0,...,0 & 0,...,0 & 0,...,0 & 0,...,0 & 
                                     {1,...,1} & {1,...,1} \\
\end{tabular}
\nonumber
\eeqn
}}

It should be emphasized that the success of free fermionic models
in providing a viable framework for reproducing the low energy
phenomenology makes evident the need for better understanding
of this class of models, both from the phenomenological
point of view, as well as trying to understand the
nonperturbative mechanism which fixes the string vacuum.
It should be further emphasized that because the
free fermionic construction is formulated at an
enhanced symmetry point in the string moduli space,
it is very natural to expect that the true
string vacuum should indeed be found in the
vicinity of this point. The structure of the 
$Z_2\times Z_2$ orbifold, which underlies the free
fermionic models, then seems particularly suited
for constructing three generation models.
It is a very intriguing fact that precisely
where one would have expected to find the
true string vacuum, indeed the most realistic
string models have been found. The further
success of the free fermionic models in producing
models with solely the MSSM charged spectrum
in the observable sector then provides further
evidence for the assertion that the true string
vacuum is indeed located in this vicinity.
Elaborate exploration of the realistic free fermionic
models is therefore vital.

The detailed massless spectrum of the FNY model
and quantum charges are given in refs. \cite{fny,cfn2}.
Here we briefly recap the notation used in this paper. 
The massless spectrum includes
three generations from the sectors ${\mb}_1$, ${\mb}_2$ and ${\mb}_3$. 
The Neveu--Schwarz (NS) sector produces the gravity and gauge multiplets,
three pairs of  electroweak doublets 
$\{h_1, h_2, h_3, {\bar h}_1, {\bar h}_2, {\bar h}_3\}$,
seven pairs of $SO(10)$ singlets with observable $U(1)$ charges, 
$\{\p{12},\pb{12},
   \p{23},\pb{23},
   \p{13},\pb{13},
   \p{56},\pb{56},
   \pp{56},\pbp{56},
   \p{4},\pb{4},
   \pp{4},\pbp{4} \}$,
and three scalars that are singlets
of the entire four dimensional gauge group, $\{\p{1},\p{2},\p{3}\}$.
The states from the NS sector carry vector--like charges with respect to
all unbroken $U(1)$ symmetries.
The states from the sectors which are combinations of  
$\{{\bf1},{\mb}_{1,2,3,4},\malpha\}+2\mbeta$ are generically denoted by
$V_n$, $n=1,2,\cdots$.
The states from the sectors with some combination of
$\{{\bf 1},{\mb}_{1,2,3,4},\malpha\}$ + $\mbeta$
are generically denoted by $H_n$, $n=1,2,\cdots$.
A superscript ``$s$'' denotes when a respective $H$ or $V$ field
carries only $U(1)$ charges and 
is a singlet for each of the non-Abelian gauge groups.
The $V_n$ and $H_n$ states are vector--like with respect to some
$U(1)$ currents but can be chiral with respect to others.

\section{Flat MSSM directions of the FNY model }

\subsection{Spacetime supersymmetry and $D$-- \& $F$--constraints}

In supersymmetric models, each chiral spin--$\half$ $\psi_m$ fermion
is paired with a scalar field $\varphi_{m}$ to form a
superfield $\Phi_{m}$.
The potential $V(\varphi)$ for the scalar fields 
receives contributions from $D$--terms 
\beqn
D_a^{\alpha}&\equiv& \sum_m \varphi_{m}^{\dagger} T^{\alpha}_a \varphi_m\,\, , \label{dtgen} 
\eeqn
where $T^{\alpha}_a$ is a matrix generator of the gauge group $g_{\alpha}$ 
for the representation $\varphi_m$,  
and from $F$--terms, 
\beqn
F_{\p{m}}   &\equiv& \frac{\partial W}{\partial \Phi_{m}} \label{ftgen}\,\, . 
\eeqn
The scalar potential has the form
\beqn
 V(\varphi) = \half \sum_{\alpha, a} g_{\alpha} D_a^{\alpha} D_a^{\alpha} +
                    \sum_m | F_{\p{m}} |^2\,\, .
\label{vdef}
\eeqn

For an Abelian group, $U(1)_i$, (\ref{dtgen}) reduces to
\beqn
D^{i}&\equiv& \sum_m  Q^{(i)}_m | \varphi_m |^2 \label{dtab}\,\,  
\label{diab}
\eeqn
where $Q^{(i)}_m$ is the $U(1)_i$ charge of $\varphi_m$.  
When one Abelian group $U(1)_A$ is anomalous\footnote{If initially the anomaly
is contained in two or more $U(1)_{A,i}$,
then the anomaly can always be rotated into a single $U(1)_A$
by a unique rotation.}, 
i.e., when the trace over the massless fields of its charge 
is non--zero, 
\beqn
\Tr Q^{(A)}\ne 0\,\, ,
\label{qtnz}
\eeqn 
then (\ref{dtab}) is modified by the appearance of an additional term
$\eps$ on the right--hand side,  where
\beqn
\eps &\equiv& \frac{g^2_s M_P^2}{192\pi^2}\Tr Q^{(A)}\, .
\label{fidt}
\eeqn
Here $g_{s}$ is the string coupling and $M_P$ is the reduced Planck mass, 
$M_P\equiv M_{Planck}/\sqrt{8 \pi}\approx 2.4\times 10^{18}$. 
The FI $D$--term $\eps$ results from the standard string theory anomaly 
cancellation mechanism \cite{u1a}. The universal Green--Schwarz relations,
which result from modular invariance constraints,
remove all Abelian triangle anomalies except those involving
either one or three $U_A$ gauge bosons. 
The string anomaly cancellation mechanism breaks $U_A$ and, 
in the process, generates $\eps$.\footnote{The 
form of the FI $D$--term was determined from string theory assumptions.
Therefore, a more encompassing $M$--theory \cite{mth} might
suggest modifications to this FI $D$--term. 
However, recently it was argued that $M$--theory
does not appear to alter the form of the FI $D$--term \cite{jmr}. 
Instead an $M$--theory FI--term should remain identical to the FI--term 
obtained for a weakly--coupled
$E_8\times E_8$ heterotic string, independent of the size of 
$M$--theory's 11$^{\rm th}$ dimension.}

Spacetime supersymmetry is broken when the scalar potential
acquires a positive--definite VEV. 
Thus, the FI $D_A$--term $\eps$
breaks supersymmetry near the string scale, with
\beqn 
V \sim g_{s}^{2} \eps^2\,\, ,\label{veps}
\eeqn  
unless a set of scalar VEVs, $\{\vev{\varphi_{m'}}\}$, 
carrying anomalous charges $Q^{(A)}_{m'}$ can cancel $\eps$ by
making a contribution to $D^{(A)}$ of equal magnitude but opposite sign:
\beq
\vev{D^{A}}= \sum_{m'} Q^{(A)}_{m'} |\vev{\varphi_{m'}}|^2 
+ \eps  = 0\,\, .
\label{daf}
\eeq
Further, maintaining supersymmetry also requires any set of scalar VEVs 
satisfying eq.\ (\ref{daf})
to also be $D$--flat 
for all of the non--anomalous Abelian and non-Abelian gauge groups as well, 
\beq
\vev{D^{(i,\alpha)}}= 0\,\, . 
\label{dana}
\eeq

The appearance of a given superfield $\Phi_{m}$ in the superpotential $W$ 
imposes additional constraints on flat directions via the associated
 $F$--term, (\ref{ftgen}).
$F$--flatness (and thereby supersymmetry) can be broken through an 
$n^{\rm th}$ order 
$W$ term containing $\Phi_{m}$ when all of the additional fields in the term acquire VEVs,
\beqn
\vev{F_{\p{m}}}     &\sim& \vev{{\frac{\partial W}{\partial \Phi_{m}}}} \label{fwnb1}\\
                &\sim& \lambda_n \vev{\varphi}^2 (\frac{\vev{\varphi}}{\MS})^{n-3}\,\, ,
\label{fwnb2}
\eeqn
where $\vev{\varphi}$ denotes a generic scalar VEV.
If $\Phi_{m}$ also takes on a VEV, then
supersymmetry can be broken simply by $\vev{W} \ne 0$. 
Therefore we also demand that $\vev{W} = 0$ for each {\it individual} term.

The higher the order, $n$, of an $F$--breaking term, 
the stronger the suppression of the supersymmetry breaking scale 
below the string scale. 
This scale suppression normally results from a 
product of two effects:
(i) a factor  $\sim (\frac{\vev{\varphi}}{\MS})^n < 1$ 
(where typically $\vev{\varphi}\sim \frac{1}{10}\MS$ in weakly coupled models)
and 
(ii) less than ``factorizable'' growth, with increasing $n$,
of the world-sheet correlation function integral $I_{n-3}$, which is contained 
within the non--normalizable $(n>3)$
coupling constants, $\lambda_n $. 
By this we mean that $I_{n-3}\ll (I_1)^{n-3}$ for $n>4$.  
In the weakly coupled case, 
$F$--breaking terms with orders as high as $n\sim 17$ can generate  
supersymmetry breaking at an energy scale too far above the electroweak
scale. Note that for our flat directions, each
$<\varphi>/\MS$ contributes a suppression factor of approximately $1/30$ to 
(\ref{fwnb2}).  

As the string coupling 
increases (and the physics becomes less perturbative
in nature)
the {\it string scale} can be significantly lowered.
For strong coupling, 
the mass scale in the denominator of (\ref{fwnb2})  should be
replaced by the string scale $M_{str}$.
The diminished suppression from each $<\varphi>/M_{str}$ factor
then requires that $F$--flatness be maintained to an even higher order.  

$F$--flatness can be broken by two classes of 
superpotential terms, those composed of: 
(i) only the VEV'd fields 
and 
(ii) the VEV'd fields, and a single field without a 
VEV.\footnote{The first class of terms was referred to as ``type A'' 
and the second as ``type B'' in \cite{cceel2}.} 
Obviously, $F$--flatness is guaranteed to a specific order $n$ in $W$
when neither class of terms appears at order $n$ or below.
In \cite{cfn2}, the three all--order flat directions,
and the one $12^{\rm th}$--order flat direction satisfying 
our ``exotic decoupling'' requirements were found 
by this technique.
While lack of the appearance of either class of term  
for a given $D$--flat direction is sufficient to guarantee $F$--flatness,
this requirement is not necessary.
The non--presence of such terms can be relaxed. 
Several terms can appear without breaking $F$--flatness, provided
that the sum over all the terms
in each $\vev{F_{\p{m}}}$, and in $\vev{W}$, vanishes. 

\subsection{MSSM flat directions}

Tables I and II in Appendix A form a review of the 
the four basic classes of VEV directions (denoted FD1, FD2,
FD3, and FD4) presented in \cite{cfn2}.
These directions sustain the MSSM gauge group and produce,
in the low energy effective field theory, 
exactly the three standard 
generations of MSSM matter fields and a single set
of Higgs $h$ and $\bar{h}$ fields
as the only MSSM--charged fields. 
All other MSSM exotics are
decoupled, acquiring FI--scale masses.
FD1 is the ``root'' direction contained within the three others.
That is, all of the fields,
$\p{12}$, $\p{23}$,
$\p{4}$, $\pp{4}$, $\pb{4}$, $\ppb{4}$,
$\pb{56}$, $\Hs{15}$, $\Hs{30}$, $\Hs{31}$, and $\Hs{38}$,
that take on VEVs in FD1, do so likewise in FD2, FD3, and FD4.
FD2, FD3, and FD4 each contain one additional field with a VEV:
$\ppb{56}$, $\Hs{19}$, and $\Hs{20}$, respectively. 
FD1, FD2, and FD3 are flat to all order, while 
$F$--flatness in FD4 is broken at twelfth order.
The Fayet--Iliopoulos scales (that is, the overall scales of the VEVs) 
for these four flat directions are approximately 
$6.7\times 10^{16}$ GeV, $3.9\times 10^{16}$ GeV, $4.8\times 10^{16}$ GeV,
and $6.7\times 10^{16}$ GeV, respectively.

The above charged $\Phi$ fields are all vector--like for 
{\it all} Abelian groups, 
while the corresponding $H$ and $V$ fields are not. 
Thus, a possible variation of this class of flat directions
is to allow both a $\Phi$ field and its vector partner $\bar{\Phi}$
to each take on a VEV. For example, in FD1 both  
$\p{12}$ and $\pb{12}$ can acquire VEVs, so long as
\beqn
|\vev{\p{12}}|^2 - |\vev{\pb{12}}|^2 = 3 \vev{\alpha}^2,
\label{p12v}
\eeqn     
where $\alpha$ is an overall scale factor for a given flat direction
and is specified in Table II of Appendix A.
Indeed, we examine flat directions varieties wherein 
$\pb{12}$ picks up a VEV in each of FD1 through FD4; 
we refer to these respective modified directions as FD1V through 
FD4V.\footnote{FD1V is therefore embedded in FD2V, FD3V, and FD4V.} 
For FD2 we also consider the case where $\pp{56}$ receives a VEV alongside 
$\ppb{56}$. This particular variation is denoted FD2$^{'}$.   
Thus, FD2$^{'}$V signifies the presence of VEVs for both 
$\pb{12}$ and $\pp{56}$.
Generally, the appearance of VEVs for both fields in a 
vector--pair results in new flatness constraints with often 
important phenomenological implications. 
We discuss vector--pair constraints in Subsection 4.3.

\section{Phenomenology of flat directions}

Phenomenology of a model is generally
vastly altered 
by the turning on of a  
flat direction \cite{cfn1,cfn2}. Further, the specific 
phenomenological modifications from that of the ``unVEVed''
model can vary substantially for different directions.
We survey the respective phenomenological features
of our three all--order and our one twelfth order
singlet flat directions (the latter considered primarily for comparison)
presented in \cite{cfn2} for the FNY model.
We investigate, in particular, the MSSM three generation mass hierarchies 
and the effective $\mu$ term(s) for each direction.  

\subsection{Field content and decoupling of FI--scale massive fields}

The functions of our singlet flat directions were twofold.
In addition to cancelling the FI--term, we required each
direction to decouple all exotic fields carrying fractional 
electric charge
(that is, one $SU(3)_C$ triplet/antitriplet pair,
four $SU(2)_L$ doublets, 16 NA singlets,  
two $SU(2)_H$ doublets, and two $SU(2)_{H^\prime}$ doublets) 
from the effective low energy field theory.
The fractional fields, denoted in (\ref{ghe}) as $E^{'}_i$, are decoupled 
from the low energy effective field theory of each flat direction 
through a generalized Higgs effect. (See Appendix B.)
That is, all of the fractional fields appear in various (relatively low) 
$n^{\rm th}$--order 
superpotential terms containing $n-2$ flat direction field VEVs $<X_j>$,
\beq
E^{'}_{i_1} E^{'}_{i_2} <X_{j_1}> \frac{<X_{j_2} X_{j_3}\cdots X_{j_{n-2}}>}{(\MS)^{n-3}}\, .  
\label{ghe}
\eeq

The flat direction VEVs $\vev{X_i}$ are generally constrained
to be around the FI scale, $\vev{\alpha}$.\footnote{An exception to this
is the overall scale of the four $\p{4}$--related VEVs, which 
is unconstrained, provided the difference of the norms of these
fields is around the FI scale. The overall scale for VEVs of vector--like
pairs of fields provides a similar exception.} 
Thus, masses for the fields that are decoupled through 
renormalizable (i.e., $n=3$) 
terms are also expected to be around the FI scale, 
while masses of fields decoupled through non--renormalizable ($n>3$) 
order terms are suppressed below the FI scale 
by factors of $\sim (\vev{\alpha}/\MS)^{n-3}$.
{\it All of the fractionally charged exotics decouple through renormalizable
terms, except for the $SU(3)_C$ vector--like triplet pair.} The triplets
appear in a fifth order mass term \cite{cfn1}. 
Thus, we expect the vector triplet to receive  a mass
smaller, perhaps, by a factor of around $1/10$ to $1/100$.

Each of our four flat directions decouples not only 
the fractionally--charged fields,
but also the remaining MSSM--charged exotics with integer electric charge.
Like their fractionally charged counterparts, 
the six integer charged $SU(2)_L$ doublets receive mass through third order 
terms.
Various sets of additional NA singlets and NA 
hidden sector fields are also decoupled by the flat directions.

The additional fields decoupled by each direction,
along with the superpotential terms responsible for the decoupling, 
are also specified in Appendix B. 
The massless and the massive (decoupled) superfield content 
and related superpotential terms
are first presented for FD1 in this appendix. Following this, 
the alterations 
to mass eigenstates resulting from replacement of FD1 
by each of the other VEV directions are listed. 
In our ``root direction'' FD1, 
15 of the 44 electrically uncharged $U(1)$--charged NA singlets
and one of the totally uncharged fields, $\p{1}$, 
receive masses from third order terms;
four more singlets receive mass at fifth order;
eight of the 30 NA hidden sector 
states with no electric charge
become massive from third order terms, 
four such fields from fourth order terms, and four from fifth order terms. 
Several fields remain massless through at least seventh order;
21 of the $U(1)$--charged NA singlets,
(including the right--handed neutrino singlets, $\Nc{1,2,3}$),
two uncharged singlets, $\p{2,3}$, 
one $SU(3)_H$ triplet/anti--triplet pair, six $SU(2)_H$ doublets, and
six $SU(2)_{H^\prime}$ doublets.

The renormalizable term $\ppb{56} \Hs{19} \Hs{20}$
prevents more than one of its three associated fields from taking on a VEV 
simultaneously. Hence the distinctions in VEVs between FD2, FD3,  and FD4.
In FD2 (and all of its variations), $\Hs{19}$ and $\Hs{20}$ become FI--scale 
massive, 
while in FD3(V) the corresponding massive states are 
$\ppb{56}$ and $\Hs{20}$, 
and in FD4(V) they are $\ppb{56}$ and $\Hs{20}$. 
FD3 and FD3V also give FI--scale mass both to another $SU(2)_H$ 
doublet and to another $SU(2)_{H^\prime}$ doublet.
A novel aspect of the FD4 and FD4V classes, perhaps identifying them from
the rest, is their rendering of a near FI--scale Majorana mass to  
a right--handed neutrino singlet, $\Nc{1}$, and to the associated singlet field
$\Vs{31}$ through a seventh order term,
\beqn
<\pp{4} \Hs{15} \Hs{20} \Hs{30} \Hs{31}> \Nc{1} \Vs{31}\,\, .
\label{ncsot}
\eeqn     

A few other eigenstate differences between FD1 and others are also
especially significant.
For example,
in all ``non--V'' directions (that is FD1, FD2$^{(')}$, FD3, and FD4) 
wherein $<\pb{12}>= 0$, 
$\pb{12}$ is a massive eigenstate and $\p{23}$ is a massless eigenstate. 
However, in the ``V'' directions, FD1V, FD2$^{(')}$V, FD3V, and FD4V,  
$\pb{12}$ receives a VEV.
This triggers the rotation of the 
the $\pb{12}$ and $\p{23}$ eigenstates into the massless state
\beqn
\pb{12}^{(1)}\equiv \frac{1}{\sqrt{|\vev{\pb{12}}|^2 + |\vev{\p{23}}|^2}}
(\vev{\pb{12}}\pb{12} - \vev{\p{23}}\p{23})\,\, , \label{pb1o}
\eeqn
and into the orthogonal FI--scale massive eigenstate,
\beqn
\pb{12}^{M}\equiv \frac{1}{\sqrt{|\vev{\pb{12}}|^2 + |\vev{\p{23}}|^2}}
(\vev{\p{23}}\pb{12} + \vev{\pb{12}}\p{23})\,\, . \label{pb1ob}
\eeqn

Acquisition of a VEV by $\pb{12}$ also transforms the massless Higgs 
eigenstate from simply  
\beqn
\hbo &\equiv& \hb{1} \label{hba}\\
     &{\rm into}& \non\\
\hbo &\equiv& \frac{1}{\sqrt{|\vev{\pb{12}}|^2 + |\vev{\Hs{31}}|^2}} 
           (\vev{\Hs{31}} \hb{1} - \vev{\pb{12}} \hb{4})\,\, . 
\label{hba2}
\eeqn
In Section 4.3 we will see that the scale of $\vev{\pb{12}}$, undetermined
perturbatively, is extremely relevant to both the inter-- 
and intra--generational mass hierarchy.

Another important phenomenological effect appears in 
the FD2$^{'}$ and FD2$^{'}V$ directions.
In all but these two directions, 
$\pp{56}$ is massive while $\Hs{15}$ is massless.
However, the VEV of $\pp{56}$ in FD2$^{'}$ and FD2$^{'}V$, rotates these
fields into the massless eigenstate,
\beqn
\pp{56}^{(1)}\equiv\frac{1}{\sqrt{|\vev{\pp{56}}|^2 + |\vev{\Hs{15}}|^2}}(
\vev{\pp{56}}\pp{56} - \vev{\Hs{15}}\Hs{15})\,\, . \label{pbm} 
\eeqn
and into the orthogonal massive eigenstate
\beqn
\pp{56}^{M}\equiv\frac{1}{\sqrt{|\vev{\pp{56}}|^2 + |\vev{\Hs{15}}|^2}}(
\vev{\Hs{15}}\pp{56} + \vev{\pp{56}}\Hs{15})\,\, . \label{pbm2} 
\eeqn
The VEV of $\pp{56}$ also carries important hierarchical implications.

\subsection{Superpotentials}

We have computed the effective low energy superpotentials, denoted herein as 
$W^{\rm FD1}$, $W^{\rm FD2}$, $W^{\rm FD3}$, $W^{\rm FD4}$,
$W^{\rm FD1V}$, $W^{\rm FD2V}$, $W^{\rm FD3V}$, $W^{\rm FD4V}$,
$W^{{\rm FD2}^{'}}$, and $W^{{\rm FD2}^{'}V}$, 
for each of the four singlet directions FD1, FD2, FD3, FD4, and six of their 
vector pair variations, respectively.  
We present each superpotential up through sixth order in Appendix C.
The superpotential for FD1 is listed in total. For the remaining
flat directions, only the additional terms not present in $W^{\rm FD1}$
are given.
The terms in each superpotential are categorized according to whether they 
contain, in addition to singlet fields: 
(i)  nothing else, 
(ii) only standard MSSM fields, 
and/or right--handed neutrinos,  
(iii) both MSSM fields and hidden sector 
non--Abelian fields, or
(iv) only hidden sector non--Abelian fields.

The terms in each of our 10 low energy effective superpotentials
are those originating from seventh or lower order terms in 
the original ``un--VEV'ed'' superpotential, $W^o$.
A specific term of order $n_{o}$ in $W^o$ generates
for a given flat direction, a term of  
effective order $n_{e} = n_{o} - n_v$, when $n_v$ of the fields in the 
$W^o$ term acquire VEVs. 
Since we expect the effective coupling constant coming from an
$n_{o}$--order term in $W^o$ to be smaller than that of  
a comparable $n_{e}^{\rm th}$--order $W^o$ term when $n_v>0$
(except perhaps when $n_v=1$ and $n_e=3$),
we refer to such a term as
a ``suppressed'' effective $n_{e}^{\rm th}$ order term in the flat direction's
effective superpotential.   

To remove some trivial redundancies,
the terms 
we list in the various effective superpotentials 
are only those for which the maximal number of possible VEVs are realized. 
For each such term, the existence in the same superpotential 
of related higher order terms containing  fewer VEVs is implied.
Consider a specific effective third order MSSM superpotential term 
for FD1:
$<\pp{4} \Hs{15} \Hs{30} \Hs{31}> \Nc{1} \Hs{20} \Vs{31}$. This term
implies the presence of a quartet of related effective fourth order terms,
including for example,
$<\Hs{15} \Hs{30} \Hs{31}> \Nc{1} \pp{4} \Hs{20} \Vs{31}$, several
effective fifth and sixth order terms,
and the seventh order one,  
$\Nc{1} \pp{4} \Hs{15} \Hs{20} \Hs{30} \Hs{31} \Vs{31}$.\footnote{A 
term containing a VEV of an FI--scale massive field
does not imply higher order terms wherein the VEV is replaced by the field.}

For each flat direction, the decoupling of the associated FI--scale 
massive fields 
drastically simplifies the corresponding superpotential 
in comparison to $W^{o}$.
Only a  handful of third through sixth order $W^{o}$ terms  
survive field decoupling for any direction. 
Large numbers of $W^{o}$ terms do not appear until the seventh order. 
Relatedly, seventh order $W^o$ terms are the primary source of the 
variations among our ten low energy effective superpotentials. 
It it for these reasons that we include up to 
seventh order $W^{o}$ contributions to those ten superpotentials.  

The small number of surviving third through six order $W^{o}$ terms results in
a division between singlet, MSSM, and NA hidden sector terms at low order.
For example, the only mixed MSSM--hidden sector terms  
(henceforth, simply referred to as ``mixed terms'') in $W^{FD1}$,
besides the many with seventh order $W^{o}$ origin, 
are four effective fifth order terms appearing at sixth order in $W^{o}$. 
Similarly, besides gaining four additional mixed terms from seventh order 
in $W^{o}$, 
FD1V only acquires two new effective fourth order terms.
The two latter terms have their origin at  
fifth and sixth order in $W^{o}$.
FD2$^{'}$V merely results in an additional mixed effective fifth order term 
with seventh order origin, while FD3 leads to three similar terms.  
FD2, FD2$^{'}$, and FD2V, FD3V, FD4, and FD4V generate no additional mixed 
terms.

Effective low order, especially renormalizable, terms 
with unsuppressed couplings are few in number.  
For example, in the FD1 superpotential, $W^{FD1}$,
the singlet sector content   
of $W^{FD1}$ is just two unsuppressed renormalizable terms. 
FD1V and FD2$^{'}$V offer a sole additional singlet term, while 
the remaining flat directions provide for no others at all.  

A general property of the flat direction superpotentials is, indeed, 
that their more distinguishing terms owe their origin to seventh (or higher) 
order in $W^{o}$. 
This implies a type of ``strong stability'' for an MSSM field theory 
realized from the FNY model. 
That is, the MSSM low energy effective field theory is quite robust
against variation of the \MSSSM generating VEVs
that are flat to at least $12^{\rm th}$ order.

\subsection{MSSM three generation mass matrices}

Viable three generation mass hierarchies present very
strong constraints that a realistic model should satisfy.
Whether or not the FNY model in particular can face this
challenge remains to be seen. 
Even models in the close neighborhood of a candidate 
solution may likely not be able to satisfy all mass
criteria. Here we investigate the structure of
the three generation 
up, down, electron, and neutrino mass matrices
for each of our singlet flat solutions.
First we will consider the up, down, and electron
matrices and then separately discuss the 
Dirac and Majorana matrices for neutrinos.

\subsubsection{Quark and electron masses}

The root of all of our flat directions,
FD1, presents a good start to viable quark and electron 
mass matrices. When respective components $\hb{1}$ and $\h{3}$
of the Higgs pair $\bh$ and $h$ 
acquire VEVs, FD1 gives
an unsuppressed renormalizable mass to exactly one
generation, via the terms
\beqn
    g \hb{1} [ \Q{1} \uc{1} + \L{1} \Nc{1} ] + 
    g \h{3}  [ \Q{3} \dc{3} + \L{3} \ec{3} ], 
\label{effw3oa}
\eeqn
where $g\equiv g_s \sqrt{2}$ is the physical four--dimensional 
gauge coupling constant.
These four terms are, in fact,
the only surviving renormalizable $W^{o}$ 
MSSM--class terms. 
An interesting property of this
model is that the top and bottom quarks (or at least their
primary components) do not come from the same
$(3,2)$ representation of $SU(3)_C\times SU(2)_L$.
Rather the top is a component of $\Q{1}$ and 
the bottom quark is a component of 
$Q_{3}$.\footnote{Our index on an MSSM field does not
correspond to generation number, but rather to
the string boundary sector
${\mathbf b}_1$, ${\mathbf b}_2$, or ${\mathbf b}_3$
from which the field originates.} 
From eq.\ (\ref{effw3oa}) we see that the phenomenologically
successful relation \cite{buras}, $m_b = m_{\tau}$ at the unification
scale, is maintained. Note also that in this model all
the heavy generation Yukawa couplings are obtained at the
cubic level of the superpotential, which differs from the
case in some other free fermionic models in which the
bottom quark and tau lepton Yukawa couplings necessarily arise from
nonrenormalizable terms \cite{eu}.
  
The additional, non--renormalizable
mass terms for the ups, downs, and
electrons resulting from each of the other flat directions 
appear in the mass matrices
$M_{u^c,Q}$, $M_{d^c,Q}$, $M_{e^c,L}$.
The components $m_{ij}$ of these matrices are defined by the convention   
\beqn 
\sum_{i,j} m_{ij} f^c_{i} F_j &=& 
 ( f^c_{1}, f^c_{2}, f^c_{3}) M_{u^c,Q} ( F_{1},  F_{2}, F_{3})^{T}\, ,
\label{mmu1}
\eeqn
where $(f^c,F)$ ranges over $(u^c, Q)$, $(d^c, Q)$, $(e^c, L)$, and
$(N^c, L)$.\footnote{The component indices, $i$ and $j$, of our mass matrices 
(\ref{maru}), (\ref{mard}-\ref{marn}) 
carry the same boundary sector interpretation
as the fields themselves. Thus, the top quark mass term appears in
position $(1,1)$ in the up--class matrix, rather than in position  
$(3,3)$.}
In these matrices $\h{1}$ and $\h{3}$ are two components of the 
physical Higgs mass eigenstate $h$. 
$\hb{1}$ and $\hb{4}\equiv \Hs{34}$ are the parallel for $\bar{h}$. 

The zero in these matrices are valid {\it to all finite orders}. 
The terms that are generated by a particular flat direction are marked
by the superscripts ${(r)}$ of the coupling constants
$\lambda^{(r)}_{n}$. The given mass term 
is also generated by all flat directions that contain the one specified. 
For instance, an $r=1V$ superscript indicates the associated mass
term appears for FD1V and for all flat direction in which FD1V is embedded.
Likewise, a term with superscript $r=2^{'}$ appears for both FD2${'}$ and
FD2${'}$V.
The coupling constant subscript $n$ indicates that
the given mass term appears at $n^{\rm th}$--order in $W^o$.

\def\ch{c_{h^{'}}}
\def\chb{c_{{\bar{h}}^{'}}}
\beqn
&&M_{\uc{i},\Q{j}}=\non\\
&&\left ( 
\begin{array}{ccc}
  \hb{1}g & 
\bhp\chb\lambda^{(1V)}_{8} \frac{\vev{\p{23}\pp{4}\Hs{15}\Hs{30}}}{\MS^5}&  
0 
\\
&&\\
\bhp\chb\lambda^{(1V)}_{7} \frac{\vev{\p{23}\Hs{15}\Hs{30}}}{\MS^4}& 
\hb{1} \lambda^{(1V)}_{12}\frac{\vev{\pb{12}\p{23}^{2}\pp{4}\Hs{15}^{2}\Hs{30}^{2}\Hs{31}}}{\MS^9}& 
0 
\\
&&\\
0& 
0&
\{\hb{4} \lambda^{(1V)}_{4} \frac{\vev{\Hs{30}}}{\MS}\\
&
&  
+\hb{1} \lambda^{(2^{'}V)}_{12}\frac{\vev{\pb{12}^{2}\p{23}\pp{56}\p{4}\Hs{15}^{2} \Hs{30}^{2}}}{\MS^9}\}
\end{array} \right)\non\\
\label{maru}
\eeqn
  
Notice that several terms in our mass matrices  
contain doublets denoted as either
$\hpx$ and $\hbp$. These are linear combinations of $SU(2)_L$ doublets
generically defined as,
\beqn h^{'}  &\equiv& \frac{1}{\ch}
(\vev{\p{12}} \h{1} + l_h\vev{\p{23}} \h{3})\quad {\rm and}
\label{hevb}\\
\bar{h}^{'} &\equiv& \frac{1}{\chb}
(\vev{\pb{12}} \hb{1} + l_{\bh}\vev{\Hs{31}} \hb{4})\, . 
\label{hbevb}
\eeqn
where 
$(\ch)^{-1} \equiv\frac{1}{\sqrt{|\vev{\p{12}}|^2 +|l_h\vev{\p{23}}|^2}}$ and 
$(\chb)^{-1}\equiv\frac{1}{\sqrt{|\vev{\pb{12}}|^2+|l_{\bh}\vev{\Hs{31}}|^2}}$.
The coefficients $l_h$  and $l_{\bh}$ are defined below.
 
For generic values of $l_h$ ($l_{\bh}$), 
$h^{'}$ (${\bh}^{'}$) is not a mass eigenstate. 
Rather, $l_h\ne 1$ ($l_{\bh}\ne 1$) results in   
both massless and FI--scale massive components for $h^{'}$ (${\bh}^{'}$).  
Only in the $l_h\rightarrow1$ 
($l_{\bh}\rightarrow1$) limit does $h^{'}$ ($\bhp$) become
an eigenstate, with an FI--scale mass. In the respective limit for each,
\beqn
h^{'}&\longrightarrow&
h^{M}\equiv \frac{1}{\sqrt{|\vev{\p{12}}|^2 + |\vev{\p{23}}|^2}}
(\vev{\p{12}}\h{1} + \vev{\p{23}}\h{3})
\label{hma}\\
{\bh}^{'}&\longrightarrow&
{\bh}^{M}\equiv \frac{1}{\sqrt{|\vev{\pb{12}}|^2 + |\vev{\Hs{31}}|^2}}
(\vev{\pb{12}}\hb{1} + \vev{\Hs{31}}\hb{4})\, .  
\label{hbma}
\eeqn
$h^{M}$ and $\bh^{M}$ are orthogonal to the massless 
eigenstates $h$ and $\bh$ given in (\ref{hev}, \ref{hbev}).

For a given mass term,
$l_h$ $(l_{\bh})$ is determined by the ratio of the  
coupling constants for the contributions from 
$\h{1}$ and $\h{3}$
($\hb{1}$ and $\hb{4}$). Because the terms in which 
$h^{'}$ and $\bhp$ appear are all seventh order or higher,
computation of the related $l_h$ and $l_{\bh}$ is extremely
non--trivial. (See Subsection 4.3.5.)
However, 
the symmetries of the world-sheet charges of the respective 
fields in both the $h^{'}$ down and electron mass terms 
and in the $\bhp$ up and neutrino mass terms
strongly suggest $l_h = 1$ and $l_{\bh} = 1$ in all cases.
If this is true (as we suspect), 
then all $h^{'}$ and ${\bh}^{'}$ terms
become decoupled from the mass matrices.

By giving a VEV to $\pb{12}$, FD1V provides for several 
additional up mass terms, but no further terms for downs or electrons.
To the up matrix (\ref{maru}) FD1V contributes: 
(i) a fourth order diagonal term in $m_{3,3}$ 
    for the second generation,
    involving the $\hb{4}$ component of $\bh$;
(ii) a twelfth order diagonal term, $m_{2,2}$,
    for the first generation, 
    involving the $\hb{1}$ component of $\bh$; and
(iii) seventh and eighth order  
      off--diagonal terms in $m_{1,2}$ and $m_{2,1}$.
The seventh and eighth order off--diagonal terms involve
${\bh}^{'}$. Only if the associated $l_{\bh} \ne 1$ will 
the respective terms contain the massless $\bh$ field and
contribute to the matrix. Should this be the case, then
(as we will argue below) 
physical constraints would most likely require
the $\hb{1}$ component of $\bh$ to strongly dominate over 
the $\hb{4}$ component. Additionally, even if  
these off--diagonal terms are non-zero,
we believe they would offer only minimal perturbations to 
the first generation mass.
Most likely, $m_{1,2} m_{2,1}/m_{1,1}\ll m_{2,2}$.
That is, relatively speaking, we estimate that
even twelfth is too ``low'' of an order 
for the first generation mass term $m_{2,2}$
to allow for a see--saw mechanism. 
Nevertheless, the fact that the dominant 
mixing would be betweeen the first and third generations is 
interesting and may have testable phenomenological implications.
FD2$^{'}$ contributes the only additional up mass term:
a twelfth order second generation term that, based on (i) above
can be ignored. Thus, the generational up mass ratios are
identical for FD1V, FD2V, FD2$^{'}$V, FD3V, FD4V. For the
remaining directions, only the top quark receives mass.   

\beqn
&&M_{\dc{i},\Q{j}}=\left ( 
\begin{array}{ccc}
0  & 0 & 0\\
&&\\
h^{'}\ch\lambda^{(2^{'})}_{7} \frac{\vev{\pp{56}\Hs{15}\Hs{31}}}{\MS^4} &
\h{1}\lambda^{(2^{'})}_{10}\frac{\vev{\p{23}\pp{56}\pp{4}\Hs{15}^2\Hs{30}\Hs{31}}}{\MS^7} & 
0\\
&&\\
0           & 0 & \h{3} g   
\end{array} \right ) \nolabel\\
&& \label{mard}
\eeqn

\beqn
&&M_{\ec{i},\L{j}}=\left ( 
\begin{array}{ccc}
0& 
h^{'}\ch\lambda^{(2^{'})}_{8} \frac{\vev{\pp{56}\pp{4}\Hs{15}\Hs{31}}}{\MS^5} 
& 0 \\
&&\\
0           &
\{\h{1} \lambda^{(2^{'})}_{10}\frac{\vev{\p{23}\pp{56}\pp{4}\Hs{15}^2\Hs{30}\Hs{31}}}{\MS^7}  
& 0\\ 
&
+\h{3} \lambda^{(2^{'}V)}_{12}\frac{\vev{\pb{12}\p{23}^2\pp{56}\pp{4}\Hs{15}^2\Hs{30}\Hs{31}}}{\MS^9}\} 
&\\ 
&&\\
0           & 0 & \h{3} g   
\end{array} \right )\nolabel\\
&&\label{mare}
\eeqn

With the exception of the trivial order FD2$^{'}$V electron mass 
(for which the tenth order FD2$^{'}$ term always dominates over)  
all first and second generation down and electron mass terms
are derived from FD2$^{'}$. Thus, the $d$ and $e$ mass
ratios are identical for FD2$^{'}$ and FD2$^{'}$V. In the other directions
only the bottom and tau fields receive mass.
If $h^{'}$ contains a massless $h$ component, then
FD2$^{'}$ produces a seventh order mass term for a second down generation 
via $m_{2,1}$ in $M_{\dc{i},\Q{j}}$ 
and an eigth order mass term for 
a second electron generation via $m_{1,2}$ in $M_{\ec{i},\L{j}}$.
In this case then
FD2$^{'}$ also ``begins'' to provide for both types of
first generation masses through
tenth order terms $m_{2,2}$ in $M_{\dc{i},\Q{j}}$ 
and $M_{\ec{i},\L{j}}$.
However, while these mass terms might suggest a first generation mass, 
they are insufficient because (if the seventh order terms exist) 
they simply rotate
the second generation $d^c$ and $e$ eigenstates, respectively.  
On the other hand, if $h^{'}$ is a massive eigenstate, then it is 
the $m_{2,2}$ term that is responsible for the 
charm and muon masses.  

Our up, down, and electron mass matrices indicate that the
``better'' phenomenology is clearly found along the flat direction
FD2$^{'}$V. Since FD2$^{'}$V contains both FD1V and FD2$^{'}$,
{\it all} of the terms in the (\ref{maru}), (\ref{mard}), and (\ref{mare}) 
appear in this direction. However, as we shall see in Section 4.4,
there is a cost for this superior phenomenology:
allowing both components of the two vector pairs $(\p{12},\pb{12})$,
and $(\ppb{56},\pp{56})$ to acquire VEVs offers new dangers 
to low order $F$--breaking. However, phenomenologically consistent solutions
to this do appear possible.  As we discuss in Section
4.3.4, part of the solution is suggested
independently by our mass matrices and the Higgs fields components.  
However, before analysis of the Higgs fields, we investigate
the neutrino mass matrices.

\subsubsection{Neutrino masses}

Let us now consider the three types of neutrino mass terms:
Majorana doublet terms $m \L{i} \L{j}$, Dirac terms $m \L{i} \Nc{j}$, 
and Majorana singlet terms $m \Nc{i}\Nc{j}$.
Simply by conservation of gauged Abelian charges, 
we can show that no Majorana doublet terms appear for any of our flat 
directions. This is favorable for a good see--saw mechanism.

As we have already shown in (\ref{effw3oa}), FD1 provides for 
a renormalizable third generation Dirac mass term for $m_{1,1}$ from
$g \hb{1} \L{1} \Nc{1}$. To this, FD1V would add three diagonal terms:
a fifth order minor perturbation to $m_{1,1}$ (containing $\hb{4}$),
a sixth order $\hb{4}$ contribution for $m_{2,2}$, and
a {\it possible} twelfth order $\bhp$ to $m_{3,3}$
(See the Dirac mass matrix (\ref{marn}) below).
Thus, FD1V would imply significant mass difference between first
and second generation neutrinos. FD2$^{'}$V would, however, alters this
through its addition of a sixth order $\hb{4}$ contribution to
$m_{3,3}$, thereby balancing the first and second generation mass scales.   
This flat direction also provides a (trivial) $14^{\rm th}$ order perturbation
to $m_{2,2}$.\footnote{Each of these FD2$^{'}$V terms  
requires a VEV for the vector--partner of $\pb{56}$.}
FD4V keeps the first and second generation mass scale distinction
by only providing trivial $16^{\rm th}$ and $22^{\rm nd}$ 
order perturbations to FD1V's $m_{2,2}$ and $m_{3,3}$ terms, respectively.  
However, should $\bhp$ have a $\bh$ component, then 
FD4 yields mixing between the third generation
and that associated with $m_{3,3}$ via 
(i) a twelfth order contribution to $m_{1,3}$ and 
(ii) a thirteenth order contribution $m_{3,1}$.
\newpage

\beqn
M_{\Nc{i},\L{j}}=\hbox to 6.0truein{\hfill}&& \non\\
\left( 
\begin{array}{ccc}
\hb{1} g&
0& 
\bhp \chb \lambda^{(4V)}_{13}\frac{\vev{X^1_{1,3}}}{\MS^{10}}\\
&&\\
&&\\
0           &
\{\hb{1}\lambda^{(2^{'}V)}_{14}\frac{\vev{X^{1a}_{2,2}}}{\MS^{11}}+  \
\bhp\chb \lambda^{(4V)}_{16}\frac{\vev{X^{1b}_{2,2}}}{\MS^{13}} 
&
0          \\ 
&
+\hb{4} \lambda^{(1V)}_{6}\frac{\vev{X^{4a}_{2,2}}}{\MS^{3}}\}
& \\ 
&&\\
\bhp\chb \lambda^{(4V)}_{12} \frac{\vev{X^1_{3,1}}}{\MS^9}&
0 &
\{\bhp \lambda^{(1V)}_{12} \frac{\vev{X^{1a}_{3,3}}}{\MS^{9}}+
  \hb{1} \lambda^{(4V)}_{22} \frac{\vev{X^{1b}_{3,3}}}{\MS^{19}}\\
  &
  &
+\hb{4}\lambda^{(2^{'}V)}_{6} \frac{\vev{X^{4b}_{3,3}}}{\MS^3}\}
\end{array} \right )&&\,\, \non\\
&&\label{marn}
\eeqn

\no where,

$X^4_{1,1}\equiv\p{12} \Hs{31}$,

$X^1_{1,3}\equiv \pb{12}\p{23}(\p{4}^2+{\pp{4}}^2)\Hs{15}^2 \Hs{20} \Hs{30} \Hs{31} \Hs{38}$,


$X^{1a}_{2,2}\equiv\pb{12}^2\p{23}^2\p{4}^{(')}\p{56}\pp{56}{\Hs{15}}^2
{\Hs{30}}^2$,

$X^{1b}_{2,2}\equiv\pb{12}\p{23}^2[\p{4}^2+\pp{4}^2]\Hs{15}^2\Hs{20}^2\Hs{30}\Hs{31}\Hs{38}^2$,

$X^{4}_{2,2}\equiv \p{23}\p{56}\H{30}$,


$X^1_{3,1}\equiv\pb{12} \p{23} \pp{4} \Hs{15}^2 \Hs{20} \Hs{30} \Hs{31} \Hs{38}$,


$X^{1a}_{3,3}\equiv\pb{12}^2\p{23}\p{4}\p{56}\Hs{15}^2\Hs{30}^2$ 

$X^{1b}_{3,3}\equiv\pb{12}^2\p{23}^2(\p{4}+\pp{4})^3\Hs{15}^4\Hs{20}^2\Hs{30}^2\Hs{31}^2\Hs{38}^2$, and

$X^{4}_{3,3}\equiv \p{56} \ppb{56} \Hs{30}$.


While some Dirac terms appear for our flat directions, 
Majorana singlet terms do not.
Local $U(1)$ charge conservation forbids 
neutrino singlet Majorana mass terms of the form
\beqn
\Nc{i} \Nc{j} \prod_{\alpha,\beta} \vev{\Phi_{\alpha}}\vev{\Hs{\beta}},
\label{mmta}
\eeqn 
for any of our flat directions. 
Alternative Majorana singlet masses arising via terms  
of the form, 
\beqn
\Nc{i} S \prod_{\alpha,\beta} \vev{\Phi_{\alpha}}\vev{\Hs{\beta}},
\label{mmtb}
\eeqn 
where $S$ is a generic singlet, are likewise forbidden, {\it with
one important exception}. FD4 converts the effective trilinear
term, 
$<\pp{4} \Hs{15} \Hs{30} \Hs{31}> \Nc{1} \Hs{20} \Vs{31}$,
into the  FI--scale mass term, 
\beqn
<\pp{4} \Hs{15} \Hs{20} \Hs{30} \Hs{31}> \Nc{1} \Vs{31}.
\label{ncm}
\eeqn
Thus, a complete and viable see--saw mechanism appears
for the third generation under FD4. If they are not decoupled, 
the off--diagonal FD4V Dirac terms could then propagate this see--saw 
mechanism to another generation.   
 
The shortage of two generations of Majorana singlet masses 
for all but FD4 and FD4V, combined with the previously discussed 
missing first generation down and electron masses,
present evidence that VEVs of NA fields may be 
necessary if viable mass matrices are to be obtained in this model.
However, FD1V offers a possible way around this for neutrinos:
another field, $\Vs{32}$, might play the neutrino singlet role
if $\bhp$ should contain a massless $\bh$ component. 
\setcounter{footnote}{0}
FD1V\footnote{FD3V (which also contains (\ref{dnm1v})) 
generates a similar $\hb{4}$ term 
$\vev{\Hs{15} \Hs{19} \Hs{31}} \hb{4} \L{1} \Vs{32}$  
at sixth order, while not producing a $\hb{1}$ counterpart.}
produces the seventh order Dirac mass term, 
\beqn
\vev{\p{4} \Hs{15} \Hs{38} } \bhp\L{3} \Vs{32}\, .  
\label{dnm1v}
\eeqn
This term, along with the FD4 term,  
$<\pp{4} \Hs{15} \Hs{30} \Hs{20} \Hs{31} > \Nc{1} \Vs{31}$,
also appears in the FD4V superpotential.
Furthermore, FD1V contains an interaction term 
for $\Vs{31}$ and $\Vs{32}$:
\beqn
<\pb{12} \Hs{30}> \Hs{29} \Vs{31} \Vs{32},
\label{dnm1vb2}
\eeqn
(which also appears in FD4V).
The combination of these terms 
offers some interesting neutrino dynamics for FD4V.   

\subsubsection{Proton decay}

In the MSSM, the dangerous proton decay operators arise from 
baryon and lepton number violating superpotential terms of the form
\beqn
W &=&\phantom{+} [\eta_1 \uc{}\dc{}\dc{} 
     + \eta_2 \Q{} \dc{}\L{}   
     + \eta_3 \L{} \L{} \ec{} ]
\non\\
  && + [\lambda_1 \Q{}\Q{}\Q{}\L{} 
     + \lambda_2 \uc{}\uc{}\dc{}\ec{}]/\MS\, 
\label{dangw}
\eeqn
where generational indices are suppressed \cite{pati}.
$\eta_i$ and $\lambda_j$ represent terms of generic order
and can contain built-in suppression factors of 
$(\Nc{}/\MS)$ and/or 
$(\vev{\phi}/\MS)^n$, where 
$\vev{\phi}$ represents either an NA singlet state VEV 
or a singlet product of NA fields,\footnote{For
products of NA fields, the appropriate extra number of 
$1/\MS$ factors is implied.} 
such as a condensate of two hidden sector vector-like fields. 
Proton decay limits imply $\eta_1\eta_2\simlt 10^{-24}$ and
$\lambda_j/M({\rm GeV})\simlt 10^{-25}$ for $\Delta(B-L)= 0$ decays. 

In the FNY model, 
the VEVs of FD1 and FD1V produce several sets 
of the dangerous operators in (\ref{dangw}).
While these operators originate from terms of at least
sixth or seventh order in $W^{o}$, 
the associated suppression factors in these 
terms do not appear strong enough
to slow proton decay sufficiently. That is, the proton 
lifetime would be significantly shorter than known limits.  
We remark that local discrete symmetries which do forbid proton
decay mediating operators to all orders of non--renormalizable
terms do appear in some three generation free fermionic models \cite{pati}.
However, such a symmetry does not seem to operate in the case
of the flat directions of the FNY model. Therefore, there
would still appear the need to find a model that 
incorporates the attractive features of the FNY model, while
at the same time, incorporates such local discrete symmetries.

\subsubsection{Effective $\mu$ term for the Higgs}


In all of our flat directions 
the massless MSSM Higgs doublets have the general composition, 
\beqn h  &\equiv& \frac{1}{\sqrt{|\vev{\p{12}}|^2 + |\vev{\p{23}}|^2}} 
(\vev{\p{23}} \h{1} - \vev{\p{12}} \h{3})\quad {\rm and}
\label{hev}\\
\bar{h} &\equiv& \frac{1}{\sqrt{|\vev{\pb{12}}|^2 + |\vev{\Hs{31}}|^2}} (\vev{\Hs{31}} \hb{1} - \vev{\pb{12}} \hb{4})\, . 
\label{hbev}
\eeqn
Thus, when $\p{12}$, but not $\pb{12}$, receives a VEV 
(as in FD1, FD2, FD2$^{'}$, FD3, and FD4), then $\bar{h}$ is simply $\hb{1}$. 
However, when  $\vev{\pb{12}}\ne 0$ 
(as in FD1V, FD2V, FD2$^{'}$V, FD3V, and FD4V)  
$\bar{h}$ becomes a linear combination of $\hb{1}$ and $\hb{4}\equiv\H{34}$. 
$\vev{\pb{12}} \ne 0$ appears a problematic issue.
As we have commented, the four possible 
non--renormalizable up mass terms containing $\hb{1}$ in (\ref{maru})
(including the two $\bhp$ terms) 
all require $\vev{\pb{12}} \neq 0$. Thus, these terms actually imply
that $\bar{h}$ has a non--zero $\hb{4}$ component. 
Furthermore, all of the remaining non--renormalizable up mass terms involve 
$\hb{4}$, thereby also requiring $\vev{\pb{12}} \neq 0$. 
Hence, either all or none of the eight non--renormalizable up terms appear.

Also note that, while all of the down mass terms in (\ref{mard}) 
and electron mass terms in (\ref{mare}) 
are seventh order or higher,  
the $\vev{\hb{4}}$ up term in $m_{3,3}$ is fourth order. 
Therefore, if we assume 
that seventh (or tenth) order terms can produce viable
down and electron second and first generations mass scales, then we would
expect a fourth order $\vev{\hb{4}}$--related mass term to upset
the generational mass hierarchy.
However, this problem can be eliminated if the fourth order up mass can 
be sufficiently suppressed to be of similar magnitude to 
a seventh (tenth) order down mass.  
This is indeed possible:
since $\hb{4}$ is only a component of the eigenstate $\bar{h}$, the fourth
order mass eigenvalue contains a suppression factor,   
\beqn
\frac{<\pb{12}>}{\sqrt{|<\pb{12}>|^2 + |<\Hs{31}>|^2}}\, . \label{p12sf} 
\eeqn
Since $<\Hs{31}>$ is an FI--scale VEV, 
when $<\pb{12}>$ is several orders of magnitude below the FI--scale, 
it would indeed seem possible for (\ref{p12sf}) to suppress 
the fourth order up mass, making it 
comparable to a seventh order down mass.\footnote{Note that in this case,
the leading component of the seventh order down mass term 
should not contain a $<\pb{12}>$ factor.}

Investigation of the set of possible effective Higgs $\mu$ terms leads to 
similar conclusions regarding $\vev{\pb{12}}$.
An effective $\mu$ term for the Higgs will appear at twelfth order in the superpotential,
\beqn
\vev{\pb{12}  \p{23}  \p{4} \pb{56}  \pp{56}  \Hs{15}  \Hs{30} \Hs{31} \Hs{38}  } 
\Hs{36} \h{1} \hb{1}/(\MS)^9 \, ,   
\label{mua}
\eeqn
if the massive field $\Hs{36}$ were to acquire an appropriate intermediate scale VEV. 
However, preceding this term at ninth order is, 
\beqn 
\vev{\pb{12}  \p{23} \pp{4}  \pp{56}  \H{15}  \Hs{30}  \Hs{31} } \h{1} \hb{1} /(\MS)^6. 
\label{mub}
\eeqn
(\ref{mub}) would clearly produce too large of an effective $\mu$--term 
unless $\pb{12}$ and/or $\pp{56}$
is far below the FI--scale.
The appearance of $\pp{56}$ in all of the down and electron mass terms 
suggests that $<\pp{56}>$ should be around the FI--scale. 
Therefore, these effective $mu$ terms also imply that 
$\vev{\pb{12}}$ cannot receive an FI--scale value in a good flat 
direction. Rather, the VEV of $\pb{12}$ should be at a low to intermediate 
scale. 
Specifically, production of a phenomenologically viable effective $\mu$ term,
\beqn 
\mu \equiv \vev{\pb{12}\p{23}\pp{4}\pp{56}\H{15}\Hs{30}\Hs{31}}/(\MS)^6,    
\label{muca}
\eeqn
in the 100 GeV to 10 TeV range requires,
\beqn 
\vev{\pb{12}} 256 \left( \frac{<\alpha>}{\MS}\right)^6 
\approx {\rm 100\ GeV\ to\ 10\  TeV},    
\label{mucb}
\eeqn
where $<\alpha>\approx 3.9\times 10^{16}$ GeV 
      and $\MS \approx 2.4\times 10^{18}$. 
This predicts a range for $\vev{\pb{12}}$ of around 
$10^{10}$ to $10^{12}$ GeV.

Additional evidence against a large $\vev{\pb{12}}$ similarly 
appears in a tenth order $\mu$ term for $\h{1}\hb{4}$,
\beqn
\vev{\p{23} \pp{56} \pp{4} \H{30}  \Hs{15}^2 \H{31}^2} \h{1}\hb{4}/(\MS)^7 
\, .   
\label{muc}
\eeqn
Here the set of VEVs generating an effective $\mu$ does not contain $\pb{12}$,
but $\hb{1}$ is replaced by $\hb{4}$. Thus, we should again 
expect $\mu$ to be significantly above the EW scale unless the
$\hb{4}$ contribution to $\bar{h}$ is either zero or extremely small,
which again implies that $<\pb{12}>$ is also far below the $FI$--scale.
The same conclusions are suggested by $16^{\rm th}$ order terms from FD4:
\beqn
(\vev{\p{12}}\h{1}+ \vev{\p{23}}\h{3})
\hb{4} \vev{\p{23}\pb{56}[\p{4}^2+\pp{4}^2]\Hs{15}^2\Hs{20}^2\Hs{31}^3\Hs{38}^2}/(\MS)^{13}\, . 
\label{FD4d1}
\eeqn

\subsubsection{Non--renormalizable coupling constants}

An important stringy aspect to the generational mass hierarchy is the numeric scale of the
the coupling constants $\lambda_{n}$ in $n^{th}$--order 
non--renormalizable mass terms. 
In free fermionic models,
these coupling constant can be expressed in terms of an $n$--point 
string amplitude $A_n$.
This amplitude is proportional to a world-sheet integral $I_{n-3}$ of the correlators of the 
$n$ vertex operators
$V_i$ for the fields in the superpotential terms. \cite{nahew5,cceel,cew},
\beqn
A_n  &=&  \frac{g}{\sqrt{2}}(\sqrt{8/\pi})^{n-3}C_{n-3}I_{n-3}/(\MS)^{n-3}\,\, . 
\label{ainta}
\eeqn  
The integral has the form,
\beqn
I_{n-3}
     &=& \int d^2 z_3\cdots d^2z_{n-1}\,\,  \vev{V_1^f(\infty) V_2^f(1) V^b_3(z_3) \cdots V^b_{n-1}(z_{n-1}) V^b_{n}(0)}  
\label{intv}\\
     &=& \int d^2 z_3\cdots d^2z_{n-1}\,\, f_{n-3}(z_1=\infty,z_2=1,z_3,\cdots,z_{n-1},z_{n}=0),
\label{intf}
\eeqn
where $z_i$ is the world--sheet coordinate of the fermion (boson) vertex operator $V^f_i$ ($V^b_i$) of the $i^{\rm th}$ string state. 
$C_{n-3}$ is an ${\cal{O}}(1)$ coefficient that includes 
renormalization factors in the operator product expansion of the string vertex operators 
and target space gauge group Clebsch--Gordon coefficients.
$SL(2,C)$ invariance is used to fix the location of three of the
vertex operators at $z= z_\infty, 1, 0$. When $n_v$ of the fields 
$\prod_{i=1}^l\X_{i}$ take on VEVs, $\vev{\prod_{i=1}^l\X_{i}}$, then the
coupling constant for the effective $n_e= (n-n_v)$--th order term becomes
$A^{'}_{n_e}\equiv A_{n} \vev{\prod_{i=1}^l\X_{i}}$.   

An $n$--point string function trivially vanishes when the correlator $\vev{\prod_i V_i}$ itself
vanishes, resulting from non--conservation of at least one or more gauged
 or global (including ``Ising'') 
world-sheet charges.  When all charges are conserved, one must compute 
$I_{n-3}$ to determine the numeric value of $A_n$.
It might actually be possible for an $n$--point function 
to vanish upon integration of $\vev{\prod_i V_i}$, even when 
$\vev{\prod_i V_i}$ is non--zero     
(i.e., when all gauge, picture--changed global world-sheet, and Ising charges 
are conserved).
Typical non--zero values of $I_1$ and $I_2$ integral for 4-- and 
5--point string amplitudes
are around 100 and 340 \cite{nahew5,cceel} for free fermionic models.

As (\ref{mard}) and (\ref{mare}) indicate,
when the relevant $\h{1}$-- and $\h{3}$--term coupling constants are 
non-equal,  our second generation down and electron mass terms  
appear at seventh and eight order, respectively, 
in the superpotential.\footnote{Alternatively, when the couplings are equal, 
then actual mass terms are tenth order.} 
Thus, in the case of coupling constant {\it inequality},
an order of magnitude estimate of the mass ratio between the second and third 
generations requires knowing the coupling strength of the seventh order 
term.
This necessitates numeric computation of the associated correlation 
function integral.
To our knowledge, this has not been done to date for sixth order or beyond. 
Thus, for an order of magnitude comparison of $7^{\rm th}$ order couplings 
to $3^{\rm rd}$, (i.e. between third and second generations),
we have analyzed the integral $I_4$ associated with the superpotential term,
$\h{3}\Q{1}\dc{2}\vev{\p{23}\pp{56}\Hs{15}\Hs{30}}$.
In the zero external momentum limit we find, 
\beqn
I_4 &=& \int d^2 z_3 d^2 z_4 d^2 z_5 d^2 z_6 \non\\
    & &\frac{|z_3|^{5/4}|z_5|}{
             |\zi - 1| |\zi - z_3| |\zi-z_5|^2 |\zi - z_6| |z_3-z_4| |z_3- z_5| |z_4| }\nonumber\\
    & & \times
        \frac{\{|z_5|^2|z_6|^2/\zi
                + 2 |z_4|^2\zi\}}
            {(|1-z_3|\, |z_6|)^{3/2} (|1- z_6|\, |z_3-z_6|)^{3/4}|z_4-z_5|^4 |z_4-z_6|^2}
\label{i4a}
\eeqn
The integrand in $I_4$, computed directly from the product of 
correlations functions, initially contained many more terms than shown in 
(\ref{i4a}). However, (relating to our comments about the possibility of
an $I_{n-3}=0$ integral above),
the vast majority of terms were in fact not
holomorphic with regard to coordinates,   
but instead contained extra $(z_i - z_j)$ or 
$(\bar{z}_{i'} - \bar{z}_{j'})$ factors that could not be paired to form
$|z_i - z_j|$ factors. Since integration of such non--holomorphic terms 
results in a zero contribution to $I_4$, only the holomorphic 
(equivalently, ``non--zero'') terms are included in (\ref{i4a}).  

The second term in the integrand of $I_4$ clearly dominates over the 
first term, 
except in the $z_6 \rightarrow \infty$ limit. Thus, for a first order of 
magnitude approximation, $I_4$ reduces to,
\beqn
I_4 &=& 2 \int d^2 z_3 d^2 z_4 d^2 z_5 d^2 z_6 \non\\
    & &\frac{|z_3|^{5/4}|z_4||z_5|}{
             |\zi - z_3| |\zi-z_5|^2 |\zi - z_6| |z_3-z_4| |z_3- z_5| }\times
\non\\
    & & \times
        \frac{1}
            {(|1-z_3|\, |z_6|)^{3/2} (|1- z_6|\, |z_3-z_6|)^{3/4} |z_4-z_5|^4 |z_4-z_6|^2}\, .
\label{i4b}
\eeqn
Several poles and fixed factors of infinity (i.e. of $\zi$) are found in 
the integrand, making evaluation difficult at best.
This could indicate that the spacetime momentum should not be 
set to zero before performing the integral. 

General consideration of (\ref{i4b}) might appear to imply something
contrary to that expected of a seventh order coupling constant.
For a weakly coupled model, when $n-3$ fields acquire VEVs in an
$n^{\rm th}$ order term, we expect
the magnitude of the effective third order coupling constant 
to be far below unity, i.e., to produce strongly suppressed terms. 
However, integration over a space with infinite limits of a  
a positive--semidefinite integrand involving several poles  
makes a finite result difficult to imagine.
In this case, the resolution to obtaining a finite value 
seems to rest on the infinities in the denominator. 
The approach to computing the integral is, perhaps, not to actually integrate,
but to examine the divergent areas and expand about them.
Clearly the most highly divergent region is near $z4=z5$, but
the zeros from the $1/\zinf$ factors
might be enough to damp out the four resulting infinities.    

What value of $I_4$ is required of the seventh order down term
to satisfy the physical $m_{s}:m_{t}\approx 10^{-3}$?
The coupling constants of the third order mass terms are all $A_3= g$,
Thus, a correct order of magnitude mass ratio requires, 
\beqn
\frac{A_7 \vev{\p{23}\pp{56}\Hs{15}\Hs{30}}}{A_3}
&=&  \frac{1}{\sqrt{2}}(\sqrt{8/\pi})^{4}C_{4}I_{4}
\frac{\vev{\p{23}\pp{56}\Hs{15}\Hs{30}}}{(M_{Pl})^{4}}\non\\
&\approx& 10^{-3}\,  .
\label{ratio73a}
\eeqn
Insertion of the FD2$^{'}$ VEVs in (\ref{ratio73a}) results in
\beqn
C_4 I_4 \approx 43\, .
\label{i4v}
\eeqn
In NAHE class $Z_2\times Z_2$ models, 
$I_1$ and $I_2$ take on respective values typically around $70$ and $400$. 
Furthermore,  
$I_n$ generally increases with $n$.
Thus,  
(\ref{ratio73a}) is in disagreement with the probable magnitude
of $I_4>> 400$, given that $C_4$ is 
{\cal{0}}(1).\footnote{In strongly coupled models, 
for which substitution of $M_{string} \ll M_{Pl}$ 
in (\ref{ratio73a}) might be justified, an even smaller $C_4 I_4$ value
is found.} However, if the seventh order down mass term contains only 
the massive eigenstate $h^{m}$ (i.e. equal couplings)
then (\ref{mard}) suggests the second 
generation results from a tenth order term. The constraint on the 
associated $I_7$ becomes,
\beqn
\frac{A_{10} \vev{\p{23}\pp{4}\pp{56}\Hs{15}^2\Hs{30}\Hs{31}}}{A_3}
&=&  \frac{1}{\sqrt{2}}(\sqrt{8/\pi})^{7}C_{7}I_{7}
\frac{\vev{\p{23}\pp{4}\pp{56}\Hs{15}^2\Hs{30}\Hs{31}}}{(M_{Pl})^{7}}
\nolabel\\ 
&\approx& 10^{-3} .
\label{ratio73b}
\eeqn
This implies $I_7 \approx 200,000$. While extrapolating a pattern for
values of $I_n$ based only on the known ranges of $I_4$ and $I_5$ in FNY
is untenable, a value of $I_7$ in this range appears possible.
Thus, it is phenomenologically preferable that for the seventh order
down term $h^{'}$ reduces to the the massive $h^M$ eigenstate.

\subsection{$F$--constraints from vector pairs}

The FD2$^{'}$  and FD2$^{'}$V  variations of FD2, in which
vector partners of FD2 fields $\pbp{56}$ and
of both $\pbp{56}$ and $\p{12}$, respectively, also take on VEVs, 
result in the more phenomenologically viable mass matrices.
Note especially that if $\pp{56}$ does not acquire a VEV, then no  
new up, down, or electron mass terms are produced by the FD2 class models. 
The appearance of these vector partner VEVs does generate some new and 
dangerous (low order) non--zero $F$--terms,
\beqn
 \vev{\partial W/\partial\p{13}}& = &  g\vev{\pb{12} \p{23} }  \label{dt2}\\     
 \vev{\partial W/\partial\pb{13}}& = & 
               \lam{7}\vev{\pp{56} \pp{4} \Hs{30} \Hs{15}^2 \Hs{31}}/(\MS)^4 \label{dt3}\\     
 \vev{\partial W/\partial\Hs{16}}& = &  g\vev{\pp{56} \Hs{15}}            
\label{dt4}
\eeqn
along with a non--zero $\vev{W}$ term,
\beqn
 & \lam{8}\vev{\pb{12} \pp{56} \p{23} \pp{4} \Hs{15}^2 \Hs{30} \Hs{31}}/(\MS)^5\,\, . \label{dt1}  
\eeqn

Examination of the FNY superpotential indicates that,
when $\pbp{56}$ and $\p{12}$ acquire VEVs, 
survival of spacetime supersymmetry down to the EW scale 
requires VEVs for some NA fields as well. 
No other singlet terms can lead to  
cancellation of $F$--term components.
Particularly worrisome is (\ref{dt1}) when $\vev{\pb{12}}\ne 0$. 
Cancellation of all of the first derivatives of (\ref{dt1}) may 
necessitate that several NA fields acquire VEVs.      
However, other phenomenological issues, such as the Higgs mass scale, 
also imply that $\pb{12}$ should not acquire an FI--scale VEV. 
In the case where $\vev{\pb{12}}= 0$ or is extremely small, 
the dangerous terms reduce to (\ref{dt3}), (\ref{dt4}), and
\beqn  
 \vev{\partial W/\partial\pb{12}} & = & \lam{8}\vev{\pp{56} \p{23} \pp{4} \Hs{15}^2 \Hs{30} \Hs{31}}/(\MS)^5 + \dots 
\label{dt6}
\eeqn
Consider now (\ref{dt3}). 
Inclusion of singlet and hidden sector terms in (\ref{dt3}) through sixth order results in
\beqn  
 \partial W/\partial\Hs{16} & = & g \vev{\pp{56} \Hs{15}} +  \lam{4} \Nc{3}  \V{37}  \H{28}/\MS + \non\\ 
                            &   & \lam{5} \p{2}  \Nc{3}  \V{37}  \H{28}/(\MS)^2 + 
                                  \lam{5} \vev{\pp{4}} \Nc{3}  \V{40}  \H{28}/(\MS)^2 +\non\\ 
                            &   & \lam{6} \vev{\H{30}} \Nc{1} \Hs{29}  \V{31} \H{20}/(\MS)^3 
\label{ndt3a}
\eeqn  
A non--zero contribution to $\vev{\partial W/\partial\Hs{16}}$ from  $\p{2}  \Nc{3}  \V{37}  \H{28}$
implies a larger contribution from $\Nc{3}  \V{37}  \H{28}$.
We should also expect the suppression factors for the 
 $\lam{5}\vev{\pp{4}} \Nc{3}  \V{40}  \H{28}$ and $\lam{6} \vev{\H{30}} \Nc{1} \Hs{29}  \V{31} \H{20}$
terms to be too small to enable these terms to cancel $\vev{\pp{56} \Hs{15}}$.
Thus, $\Nc{3}$, $\V{37}$, and $\H{28}$ must all receive FI--scale VEVs when $\vev{\pp{56}}\ne 0$. This is a viable solution because
fourth order string couplings can actually be as 
large as third order couplings \cite{cew}. 

\section{Conclusions}

In this paper we investigated  
phenomenological issues of the four flat directions 
(and some of their variations) derived in \cite{cfn2}
for the ``FNY'' model \cite{fny,fc}. 
Using our ``stringent'' $F$--flat requirements, these were
the only directions found to be flat beyond seventh order
that decoupled all fractionally charged MSSM  
exotics from the low energy superpotential.
(It was as a serendipitous bonus that these four directions 
decoupled {\it all} MSSM--charged exotics.) 
Approximately 100 other VEV directions were found that removed
the fractionally charged fields, but $F$--flatness for these was broken
at only sixth or seventh order.

The first three of our flat directions are actually flat to
all finite order, while $F$--flatness of the fourth is 
broken at twelfth order. Variations on these flat directions,
in which both components of vector pairs of fields
were allowed to acquire VEVs, were also investigated.
The variations of $FD2$ were found to be the most 
phenomenologically viable with regard to mass matrices.

We examined, for the different flat directions, the 
inter--generational mass hierarchy for the MSSM quarks and leptons
and possible effective $\mu$ terms for the MSSM Higgs.
Reasonable, but not perfect, up, down, electron, and neutrino
hierarchy--producing mass matrices were found for particular 
variations of our flat directions. Possible additional $F$--flatness
constraints on the vector--pair flat directions were also
examined.

An interesting aspect of the 
Higgs fields was mutually suggested by the mass matrices,
the possible $\mu$ terms, and the additional $F$--flatness constraints.
In the FNY field basis given in \cite{fny}, the MSSM Higgs eigenstates
$h$ and $\bh$ each have two components. For several phenomenological
reasons,
the $SU(2)_L$ doublet $\hb{1}$ contribution to  
to $\bh$ must greatly outweigh $\hb{4}$'s.
Relatedly, it is only the $\hb{4}$ component of $\bh$,
and not the $\hb{1}$ contribution, that couples to the
second generation up field, ultimately giving mass to it.
This results in an additional cosine suppression factor 
for the charm mass term. This is a favorable feature for this 
model, since the
charm mass term originates at fourth order, 
while those for the strange and muon masses appear at seventh
and eight order, respectively. Thus, without the extra cosine
factor, this model would predict a several orders of magnitude 
mass difference between the charm and strange fields. 
 
Our investigation of singlet flat directions led to several pieces of
evidence suggesting that some non--Abelian fields must also acquire 
VEVs for the FNY model to produce viable phenomenology.
The additional $F$--terms arising from FD1V and FD2$^{'}$ especially
implied this. In particular, for FD2$^{'}$ it appears necessary 
for two $SU(2)_{H^\prime}$ doublets (along with a neutrino singlet) to also
acquire FI--scale VEVs.

In forthcoming \cite{cfn4}, we will generalize our current flat directions by permitting 
non--Abelian fields to acquire VEVs. The simplest non--Abelian 
extension to each of our
current direction is the addition of the only 
completely chargeless product of hypercharge--singlet
non--Abelian fields: $\V{5}\V{40}\V{10}\V{37}$. 
We shall also examine the additional flat
directions available when our basis set of 24 nontrivial $D$--flat singlet 
directions increases to a basis set of 53 non--trivial directions involving 
the 30 hypercharge--singlet hidden sector non--Abelian fields in the FNY model.

We conclude by reemphasizing that the success of the FNY model
in producing Minimal Superstring Standard Models should
not be regarded as suggesting that the FNY model is the
true string vacuum. Rather, it is our firm opinion
that the success of the FNY model in this regard,
together with the other properties of the free fermionic
NAHE--based models, such as the natural emergence of 
three generations, can be regarded as providing evidence
to the assertion that the true string vacuum will possess
some of the characteristics shared by this class of models.
Furthermore, the existence of highly realistic string models
near a maximally symmetric point in the moduli space,
provides further indication for the relevance of string
theory in nature. The continued development of the
techniques needed to seriously confront string models
with experimental data is therefore imperative. The
main focus of the present paper being the phenomenological
analysis of exact flat directions.

\section{Acknowledgments}
This work is supported in part
by DOE Grants No. DE--FG--0294ER40823 (AF)
and DE--FG--0395ER40917 (GC,DVN,JW).
\newpage
\appendix

\section{$D$-- and $F$--flat Directions} 

\def\ify{$\infty$ }
\def\ifw{${\infty}^{\ast}$ }

\def\y{$\ast$}
\def\my{$\bar{\ast}$}
\def\ny{${(-)}\atop{\ast}$}

\begin{flushleft}
\begin{tabular}{|r||r|l|ccccc|}
\hline 
\hline
FD$\#$ &$\#$&$\cal{O}(W)$  &$\{VEV_1\}$&\opb{56}&\opp{56}&\oHs{19}&\oHs{20}\\
       &VEVs& breaking     &           &        &        &        &       \\
\hline
 1    & 8+3  & $\infty$   &\y         &\y      &        &        &       \\ 
 2    & 9+3  & $\infty$   &\y         &\y      &\my     &        &       \\  
 3    & 9+3  & $\infty$   &\y         &\y      &        &\y      &       \\   
 4    & 9+3  & 12         &\y         &\y      &        &        &\y     \\
\hline
\hline
\end{tabular}
\end{flushleft}

\noindent Table I: Classes of $D$ \& $F$--flat NA singlet directions  
that yield an MSSM low energy effective field theory . 
\vskip .1truecm

\noindent
The four classes are defined by their respective set of 
non--Abelian singlet field VEV components. 
The class identification number appears 
in column one. Column two specifies the number of singlet VEVs,
with its first entry specifying 
the number of VEVs resulting from FI cancelation 
and its second entry indicating the number of $\Phi_4$--related fields to 
which additionally VEVs are assigned. 
Column three indicate the superpotential order at which flatness is broken.
The $*$'s in the remaining columns identify the field VEVs.
An $*$ implies a given field takes on a VEV, while a $\bar{\ast}$ 
implies the vector partner does instead.  
A $*$ below $\{VEV_1\}$ indicates that all fields in the set   
$\{\p{12},\p{23},(\p{4},\pp{4},\pb{4},\ppb{4}),
\Hs{30},\Hs{38},\Hs{15},\Hs{31}\}$ acquire VEVs. 

\def\ify{$\infty$ }
\def\ifw{${\infty}^{\ast}$ }

\def\x{$\phantom{1}$}
\def\m{$\phantom{-}$}

\begin{flushleft}
\begin{tabular}{|r||r|l|rrrrr|}
\hline 
\hline
FD$\#$  
&$\frac{Q^{(A)}}{112}$&$\vev{\alpha}$  
                 &$\{$\op{12},\op{23},\opp{(4)},\oHs{30}\oHs{38},\oHs{15},\oHs{31},$\}$&\opb{56}&\opp{56}&\oHs{19}&\oHs{20}\\
\hline
\hline
 1  &  -2 & $6.7\times 10^{16}$ GeV 
    &{\x}3, {\x}1, {\x}1, {\x}3, {\x}2, {\x}2, {\x}1$\phantom{\}}$
    &{\m}1  &{\m}0 &0  &0  \\  
 2  &  -6 & $3.9\times 10^{16}$ GeV
    &   10, {\x}2, {\x}2, {\x}8, {\x}6, {\x}4, {\x}2$\phantom{\}}$         
    &{\m}3  &   -1 &0  &0  \\  
 3  &  -4 & $4.8\times 10^{16}$ GeV 
    &{\x}4, {\x}3, {\x}2, {\x}8, {\x}2, {\x}6, {\x}2$\phantom{\}}$  
    &{\m}1  &{\m}0 &2  &0  \\  
 4  &  -2 & $6.7\times 10^{16}$ GeV
    &{\x}3, {\x}2, {\x}3, {\x}3, {\x}4, {\x}4, {\x}3$\phantom{\}}$  
    &{\m}2  &{\m}0 &0  &2  \\  
\hline
\hline
\end{tabular}
\end{flushleft}

\noindent Table II: 
Examples of flat directions from the four classes presented in Table I.
\vskip .1truecm

\no Column one indicates the class to which an example direction belongs.
Column two contains the anomalous charges $Q^{(A)}/112$ of the example flat directions.
Column three specifies the overall common scale of the associated VEVs.
The remaining column entries specify the ratios of the norms of the VEVs. 
$\{\p{12},\p{23},(\p{4}),\Hs{30}\Hs{38},\Hs{15},\Hs{31},\}\equiv\{ VEV_1 \}$ 
form the set of fields possessing VEVs in all four directions.
The third component VEV involves all of the $\p{4}$--related states and is 
the net value of
$\mvev{\Phi_4} + \mvev{\Phi^{'}_4} - \mvev{\bar{\Phi}_4} - \mvev{\bar{\Phi}^{'}_4}$. 
E.g., a ``1'' in the $\Phi_4$ column for FD1 specifies that   

\no $\mvev{\Phi_4} + \mvev{\Phi^{'}_4} - \mvev{\bar{\Phi}_4} - \mvev{\bar{\Phi}^{'}_4} = 1\times\vev{\alpha}^2$. 
      
\hfill\vfill\eject


\section{Field content for flat directions}

\vskip 0.5truecm

\no Massless Fields for flat direction 1:
 
\begin{itemize} 
\item  3 generations of MSSM fields: 
$(Q,\, u^c,\, d^c,\, L,\, e^c,\, N^c)_{i=1,2,3}$

\item  2 MSSM Higgs: 

$h \equiv \frac{1}{\sqrt{|\vev{\p{12}}|^2 + |\vev{\p{23}}|^2}} 
           (\vev{\p{23}} \h{1} - \vev{\p{12}} \h{3})$, 

$\bar{h}\equiv \frac{1}{\sqrt{|\vev{\pb{12}}|^2 + |\vev{\Hs{31}}|^2}} 
           (\vev{\Hs{31}} \hb{1} - \vev{\pb{12}} \H{34})$ 

\item  21 Non-Abelian Singlets: \op{2}, \op{3}, \op{12}, \op{23}, 


      $\pb{23}^{(1)}\equiv \frac{1}{\sqrt{|\vev{\p{12}}|^2 + |\vev{\Hs{30}}|^2 
                                                           + |\vev{X}|^2}}
                          (\vev{\Hs{30}}\pb{23} - \vev{\p{12}}\Hs{29} 
                                                + \vev{X}\Hs{37})$, 

where $X\equiv\p{23}\pb{56}\Hs{31}\Hs{38}$

       $\p{4}^{(1)}\equiv \frac{1}{\sqrt{|\vev{\pb{4}}|^2 + |\vev{\ppb{4}}|^2}}
                              (<\pb{4}> \p{4} - <\ppb{4}> \pp{4} )$, 

       $\p{4}^{(2)}\equiv \frac{1}{\sqrt{|\vev{\p{4}}|^2 + |\vev{\pp{4}}|^2}}
                                  (<\p{4}> \pb{4}- <\pp{4}> \ppb{4} )$, 

       $\p{4}^{(3)}\equiv \frac{(<\ppb{4}> \p{4}+ <\pb{4}> \pp{4}- 
<\pp{4}>\pb{4}- <\p{4}> \ppb{4})}{\sqrt{|\vev{\p{4}}|^2 + |\vev{\pp{4}}|^2+
                                      |\vev{\pb{4}}|^2 + |\vev{\ppb{4}}|^2}}$, 

       \op{56},\opb{56}, \oppb{56},
       \oHs{15}, \oHs{19}, \oHs{20}, \oHs{30}, \oHs{31}, \oHs{36}, \oHs{38},
       \oVs{1}, \oVs{12}, \oVs{31}, \oVs{32},

 \item 2 Hidden Sector $SU(3)_H$ (Anti)-Triplets:  \oV{4}, \oV{13}

 \item 6 Hidden Sector $SU(2)_H$ Doublets: 
\oH{23}, \oH{26}, \oV{5}, \oV{15}, \oV{39}, \oV{40}
                    
 \item 6 Hidden Sector $SU(2)_{H^\prime}$ Doublets: 
\oH{25},  \oH{28}, \oV{10}, \oV{19}, \oV{35}, \oV{37}

\end{itemize} 

\newpage
\no Fields with (near) FI--scale masses for flat direction 1:

\no (i) Exotic MSSM-Charged States
\begin{itemize} 
\item  $SU(3)_C$ Triplets:  \oH{33}, \oH{40} 

\item  $SU(2)_L$ Doublets: \oh{2}, \ohb{2}, \ohb{3}

$h^{M}\equiv \frac{1}{\sqrt{|\vev{\p{12}}|^2 + |\vev{\p{23}}|^2}}
(\vev{\p{12}}\h{1} + \vev{\p{23}}\h{3})$, 

 $\hb{4}\equiv\H{34}$, $\h{4}\equiv\H{41}$, 
  \oV{45}, \oV{46}, \oV{51}, \oV{52} 

\item  Hidden Sector $SU(2)_H$ Doublets (with fractional electric charge): \oH{1}, \oH{2}

\item  Hidden Sector $SU(2)_{H^\prime}$ Doublets (with fractional electric charge): \oH{11}, \oH{13} 
\end{itemize}

\no (ii) Non-Abelian Singlets 
 



\begin{itemize} 
\item  \op{1}, \opb{12}, \op{13}, 

$\pb{13}^{M}\equiv \frac{1}{\sqrt{|\vev{\Hs{30}}|^2 + |\vev{X}|^2}} 
                    (\vev{\Hs{30}}\pb{13} +  \vev{X}\Hs{36})$

$\Hs{36}^{M}\equiv \frac{1}{\sqrt{|\vev{\Hs{30}}|^2 + |\vev{X}|^2}}
                    (\vev{X}\pb{13} -  \vev{\Hs{30}}\Hs{36})$

$\pb{23}^{M}\equiv
\frac{[\vev{\p{12}}\pb{23} + 
(\vev{\Hs{30}}+ \vev{X})^2/\vev{\Hs{30}})\Hs{29} + 
(\vev{X}\vev{\p{12}}/\vev{\Hs{30}})\Hs{37}]}{\sqrt{|\vev{\p{12}}|^2 + 
|\vev{\Hs{30}}+ \vev{X}^2/\vev{\Hs{30}}|^2+
|\vev{X}\vev{\p{12}}/\vev{\Hs{30}}|^2}}$

$\Hs{37}^{M}\equiv
\frac{1}{\sqrt{|\vev{\Hs{30}}|^2 + |\vev{X}|^2}}
(\vev{X}\pb{23} - \vev{\Hs{30}}\Hs{37})$,

$\p{4}^{M}\equiv 
\frac{(\vev{\ppb{4}}\p{4} + \vev{\pb{4}}\pp{4} + \vev{\pp{4}}\pb{4} + 
  \vev{\p{4}}\ppb{4})}{\sqrt{|\vev{\p{4}}|^2 + |\vev{\pp{4}}|^2+
              |\vev{\pb{4}}|^2 + |\vev{\ppb{4}}|^2}}$
 
 \opp{56}, 
 \oHs{3}, \oHs{4}, \oHs{5}, \oHs{6}, \oHs{7}, \oHs{8}, \oHs{9}, \oHs{10}, 
 \oHs{16}, \oHs{17}, \oHs{18}, \oHs{21}, \oHs{22}, \oHs{32}, \oHs{37}, \oHs{39}, 
 \oVs{2}, \oVs{11}, \oVs{21}, \oVs{22}, 
 \oVs{41}, \oVs{42}, \oVs{43}, \oVs{44}, \oVs{47}, \oVs{48}, \oVs{49}, \oVs{50}   
\end{itemize} 

\no (iii)  Hidden Sector Non-Abelian Fields: 

\begin{itemize} 
 \item $SU(3)_H$ (Anti)-Triplets: 
\oH{42}, \oV{14}, \oV{24}, \oV{34}, \oH{35}, \oV{3}, \oV{23}, \oV{33}    
 
 \item $SU(2)_H$ Doublets: \oV{7}, \oV{17}, \oV{25}, \oV{27}    

 \item $SU(2)_{H^\prime}$ Doublets: \oV{9}, \oV{20}, \oV{29}, \oV{30}     
\end{itemize} 

\newpage
\no Effective FI-scale tadpole in FD1 superpotential: 
\beqn
  && \p{1}[  <\p{4} \ppb{4}> +  <\pp{4} \pb{4}> ]
\label{tp1}   
\eeqn

\no Effective FI--scale mass terms in FD1 superpotential: 
\vskip .2truecm

\no $W_3$ terms:
\beqn
  &&     \p{1}[<\ppb{4}>\p{4} + <\pb{4}>\pp{4} + <\pp{4}>\pb{4} + <\p{4}>\ppb{4}]  +    \nolabel\\   
  &&   \pb{13}[ <\p{12}>\pb{23} + <\Hs{30}> \Hs{29} ] +   \nolabel\\
  && <\p{12}> [ \h{1} \hb{2} + \Hs{36} \Hs{37} + \Vs{21} \Vs{22} + \V{29} \V{30} + \V{25} \V{27} + \V{23} \V{24} ] +\nolabel\\   
  && <\p{23}> [ \pb{12} \p{13} + \h{3} \hb{2} + \V{33} \V{34} ] +                                                 \nolabel\\
  && <\p{4}>  [ \V{45} \V{46} + \H{1} \H{2} ] +                                                 
 <\pp{4}>  [ \V{51} \V{52} + \Hs{7} \Hs{8} + \Hs{9} \Hs{10} ] +                              \nolabel\\       
  && <\pb{4}> [ \Hs{3} \Hs{4} + \Hs{5} \Hs{6} + \Vs{41} \Vs{42} + \Vs{43} \Vs{44} ] +            
 <\ppb{4}> [ \Vs{47} \Vs{48} + \Vs{49} \Vs{50} + \H{11} \H{13} ] +                          \nolabel\\
  && <\pb{56}>[ \Hs{17} \Hs{18} + \Hs{21} \Hs{22} ] +                                                     \nolabel\\       
  && <\Hs{15}>\pp{56}\Hs{16} + <\Hs{30}> \pb{13} \Hs{29} + <\Hs{31}> \h{2} \hb{4} 
   + <\Hs{38}> \H{41} \hb{3}    
\label{mw3FD1}
\eeqn

\no $W_4$ terms:
\beqn
 <\Hs{15} \Hs{30}> [ \V{3} \V{14} + \V{7} \V{17} ]    
\label{mw4FD1}
\eeqn

\no $W_5$ terms:
\beqn
 &&<\p{23}   \Hs{31}   \Hs{38}> [ \hb{4} \H{41} + \H{33} \H{40} ] + 
 <\pp{4}   \Hs{15}   \Hs{30}> [ \Vs{2} \Vs{11}+ \V{9}  \V{20} ] +\nolabel\\
 &&<\pb{56}  \Hs{31}   \Hs{38}> [ \Hs{32}\Hs{39}+ \H{35} \H{42} ]    
\label{mw5FD1}
\eeqn

\no $W_6$ terms:
\beqn
 <\p{23} \pb{56} \Hs{31} \Hs{38}> \Hs{29} \Hs{36}    
\label{mw6FD1}
\eeqn

\newpage
\no Additional Fields receiving (near) FI--scale masses for other flat directions:


\begin{flushleft}
\begin{tabular}{|r||l|l|}
\hline 
\hline
fd $\#$& non--FD1 VEVs             &New mass and tadpole terms  \\
\hline
 1V   & $\pb{12}$                   &$<\pb{12}>  [\p{13} \p{23} + \h{2} \hb{1}]$,  $<\pb{12} \p{23}> \p{13}$ \\
 2    & $\pbp{56}$                  &$<\ppb{56}> \Hs{19} \Hs{20}$ \\
 2V   & $\pbp{56},\pb{12}$          &--                 \\
$2^{'}$& $\pbp{56},\pp{56}$         &$<\pp{56}> \Hs{15} \Hs{16}$, $<\pp{56} \Hs{15}> \Hs{16}$                \\
$2^{'}$V&$\pbp{56},\pp{56},\pb{12}$ &--                 \\
 3    & $\Hs{19}$                   &$<\Hs{19}> \ppb{56} \Hs{20}$ \\
      &                             &$<\Hs{19} \Hs{21}> \V{19} \V{37}$ \\
      &                             &$<\ppb{4}  \Hs{19} \Hs{21}> \V{15} \V{40} $\\
      &                             &$<\pb{56}  \Hs{15} \Hs{19}> \Hs{18} \Hs{22} $\\
 3V   & $\Hs{19},\pb{12}$           &--                 \\
 4    & $\Hs{20}$                   &$<\Hs{20}> \ppb{56} \Hs{19}$  \\
      &                             &$<\Hs{15} \Hs{20}\Hs{31} \Hs{38} > \Vs{41} \Vs{50}$ \\   
      &                             &$<\pp{4} \Hs{15} \Hs{20} \Hs{30} \Hs{31}  > \Nc{1} \Vs{31}$ \\  
 4V   & $\Hs{20},\pb{12}$           &--                 \\
\hline
\hline
\end{tabular}
\end{flushleft}

\no The first column denotes the flat direction.  
A ``V'' in this column indicates
a vector--like partner to one of the standard fields associated with a given
flat direction is also allowed a VEV. The listings in column two indicate
a flat direction's VEVs that are not present in flat direction 1.
Column three lists the VEV--induced mass terms new to the respective 
flat directions and not present in flat direction 1. If a mass term  
for a given field is already present in a standard flat direction, 
respective higher order mass terms (should they exist) are not 
necessarily listed.

\phantom{skip line}

\no Rotated massless (with ``(1)'' superscript)
and FI--scale massive (with ``M'' superscript) 
eigenstates for non--FD1 directions:

\begin{itemize}
\item FD1V (and all embeddings): 

$\pb{1}^{(1)}\equiv \frac{1}{\sqrt{|\vev{\pb{12}}|^2 + |\vev{\p{23}}|^2}}
(\vev{\pb{12}}\pb{12} - \vev{\p{23}}\p{23})$ 

$\pb{1}^{M}\equiv \frac{1}{\sqrt{|\vev{\pb{12}}|^2 + |\vev{\p{23}}|^2}}
(\vev{\p{23}}\pb{12} + \vev{\pb{12}}\p{23})$ 

\item FD2$^{'}$, FD2$^{'}$V: 

$\pp{56}^{(1)}\equiv\frac{1}{\sqrt{|\vev{\pp{56}}|^2 + |\vev{\Hs{15}}|^2}}(
\vev{\pp{56}}\pp{56} - \vev{\Hs{15}}\Hs{15})$ 

$\pp{56}^{M}\equiv\frac{1}{\sqrt{|\vev{\pp{56}}|^2 + |\vev{\Hs{15}}|^2}}(
\vev{\Hs{15}}\pp{56} + \vev{\pp{56}}\Hs{15})$ 
\end{itemize}

\newpage

\section{Superpotentials for Low--Energy Effective Field Theories
for Flat Directions} 

\vskip 0.5truecm

\no\underline{Flat direction 1 Low Energy Effective Superpotential, $W^{\rm FD1}$:}
\vskip 0.5truecm

\no {\it Singlet Terms}

\no $W_{3}$:
\beqn
  g [\ppb{56} \Hs{19} \Hs{20} + \pb{23} \Vs{31} \Vs{32}]  
\label{effw3s}
\eeqn


\vskip .2truecm
\no {\it MSSM Terms}

\no $W_{3}$:
\beqn
&&    g \hb{1} [ \Q{1} \uc{1} + \L{1} \Nc{1} ] + 
      g \h{3}  [ \Q{3} \dc{3} + \L{3} \ec{3} ] + \non\\ 
&&  <\pp{4}   \Hs{15}   \Hs{30}   \Hs{31}>  \Nc{1} \Hs{20}  \Vs{31}   
\label{effw3o}
\eeqn

\no $W_{4}$:
\beqn
&&
    <\Hs{30}+ (\p{4}\pb{4}+ \pp{4}\ppb{4})\Hs{30}> 
     [\h{1}\Q{3}\dc{3}+\h{1}\L{3}\ec{3}]\Hs{29} + \non\\ 
&&  <\p{12}+ (\p{4}\pb{4}+\pp{4}\ppb{4})\p{12}> [ 
            \Q{1}   \Q{2}   \uc{1}   \dc{2} +  
            \Q{1}   \uc{1}   \L{2}   \ec{2} +     
            \Q{1}   \uc{2}   \L{2}   \ec{1} +  
            \Q{2}   \dc{1}   \L{1}   \Nc{2} + \non\\
&&\phantom{<\p{12}+ (\p{4}\pb{4}+\pp{4}\ppb{4})\p{12}> [}
            \Q{2}   \dc{2}   \L{1}   \Nc{1} +  
            \L{1}   \L{2}    \ec{2}  \Nc{1}    
           ] + \non\\  
          &&
    <\p{12}  \pp{4}> [
                \Q{2}   \uc{1} \L{2}   \ec{1} + 
                \Q{2}   \dc{1} \L{2}   \Nc{1} + 
                \uc{1}  \uc{1} \dc{2}  \ec{2} + 
                \uc{1}  \uc{2} \dc{2}  \ec{1} + 
                \uc{2}  \dc{1} \dc{1}  \Nc{2} + 
                \uc{2}  \dc{1} \dc{2}  \Nc{1}] + \non\\
          &&
 <\p{12} \ppb{4}> [ 
               \Q{1} \Q{1} \Q{2} \L{2}  + 
               \Q{1} \Q{1} \uc{2} \dc{2} +  
               \L{1} \L{1} \ec{2} \Nc{2} ] + \non\\
          &&
 <\Hs{15}  \Hs{30}> [     
                     \Q{1}   \Q{2}  \Q{3}  \L{3} +    
                     \Q{1}   \Q{2}  \uc{3} \dc{3}+    
                     \Q{1}   \Q{3}  \uc{2} \dc{3}+    
                     \Q{1}   \uc{2} \L{3}  \ec{3}+    
                     \Q{1}   \uc{3} \L{3}  \ec{2}] + \non\\   
&&  <\p{4} \Hs{15}\Hs{38}>[\Q{2}   \dc{2}  \L{3} +
                           \L{2}   \L{3}   \ec{2}  ]\Vs{32} +
    <\pp{4}\Hs{15}\Hs{38}>[\Q{2}   \dc{3}  \L{2} + 
                           \uc{2}  \dc{2}  \dc{3}  ] \Vs{32}+ \non\\
&&        <\p{4}   \Hs{15}   \Hs{30}> [  
                              \Q{3}   \uc{1}   \L{3}   \ec{2} +                
                              \Q{3}   \uc{2}   \L{3}   \ec{1} +                
                              \uc{1}   \uc{2}   \dc{3}   \ec{3} +             
                              \uc{1}   \uc{3}   \dc{3}   \ec{2} +  
                              \uc{2}   \uc{3}   \dc{3}   \ec{1}] +   \non\\
&&        <\pp{4}   \Hs{15}   \Hs{30}> [ 
                              \Q{2}   \Q{3}    \uc{1}  \dc{3} +              
                              \Q{2}   \uc{1}   \L{3}   \ec{3} +               
                              \Q{2}   \uc{3}   \L{3}   \ec{1}] +   \non\\
&&        <\p{23}   \Hs{15}   \Hs{30}> [
                              \Q{1}   \Q{2}   \Q{2}   \L{2} +               
                              \Q{1}   \Q{2}   \uc{2}  \dc{2}  +               
                              \Q{1}   \uc{2}  \L{2}  \ec{2} ] 
\label{effw4ssm}
\eeqn

\no $W_{5}$:
\beqn
&& [1 + <\p{4} \pb{4} + \pp{4}\ppb{4}>]
    [\h{3} \Q{2} \dc{2} + \h{3} \L{2} \ec{2}] \Vs{31} \Vs{32} + \non\\    
&& <\p{12}>[ \h{3}   \Q{2}   \dc{1} \Vs{1}  \Vs{12} +
             \Q{1}   \Q{3}   \uc{1}  \dc{3} \pb{23}  +
             \Q{1}   \uc{3}  \L{3}   \ec{1} \pb{23}  +
             \Q{3}   \dc{1}  \L{1}   \Nc{3} \pb{23}  + \non\\
&&\phantom{<\p{12}>[}
             \Q{3}   \dc{3}  \L{1}   \Nc{1} \pb{23}  +
             \L{1}   \L{3}   \ec{3}  \Nc{1} \pb{23} ]+\non\\
&&<\Hs{15}> [\Q{2}   \dc{2}  \L{1}+ \L{1}   \L{2}   \ec{2} ]\Hs{19} \Vs{32} + \non\\
&&<\Hs{30}> [\Q{1}   \Q{3}   \uc{1}  \dc{3}+      
             \Q{1}   \uc{1}  \L{3}   \ec{3}+      
             \Q{1}   \uc{3}  \L{3}   \ec{1}+     
             \Q{3}   \dc{1}  \L{1}   \Nc{3}+         
             \Q{3}   \dc{3}  \L{1}   \Nc{1}+      
             \L{1}   \L{3}   \ec{3}  \Nc{1}  ] \Hs{29} + \non\\
&&<\Hs{38}> [ \hb{1}   \L{1}   \Nc{3}  \ppb{56} \Hs{20} +        
 ( \Q{1}   \dc{2}   \L{3} +  \Q{1}   \dc{3}   \L{2}) \Hs{29} \Vs{32}] + \non\\
&&<\p{12}\p{23}> [
                \Q{1}   \Q{2}   \uc{1}   \dc{2} +  
                \Q{1}   \uc{1}   \L{2}   \ec{2} +  
                \Q{1}   \uc{2}   \L{2}   \ec{1} +  
                \Q{2}   \dc{1}   \L{1}   \Nc{2} +
                \Q{2}   \dc{2}   \L{1}   \Nc{1} +  
                \L{1}   \L{2}   \Nc{1}   \ec{2}  ] \pb{23} + \non\\
&&<\p{12} \p{4}> [   
                \Q{3}   \uc{1}  \L{3}   \ec{1}  +  
                \Q{3}   \dc{1}  \L{3}   \Nc{1}  +  
                \uc{1}  \uc{1}  \dc{3}  \ec{3}  +    
                \uc{1}  \uc{3}  \dc{3}  \ec{1}  +
                \uc{3}  \dc{1}  \dc{1}  \Nc{3}  +  
                \uc{3}  \dc{1}  \dc{3}  \Nc{1}   ] \pb{23}  + \non\\
&&<\p{12}\pb{4}> [   
                \Q{1}   \Q{1}   \Q{3}   \L{3}   +  
                \Q{1}   \Q{1}   \dc{3}   \uc{3} +  
                \L{1}   \L{1}   \ec{3}   \Nc{3}  ] \pb{23}+ \non\\
&&<\p{12}\ppb{4}> \h{3}   \L{1}   \ec{2}   \Vs{1} \Vs{12} + \non\\
&&<\p{12}\pb{56}> [
                \Q{1}   \Q{2}   \uc{1}   \dc{2} +  
                \Q{1}   \uc{1}   \L{2}   \ec{2} + 
                \Q{1}   \uc{2}   \L{2}   \ec{1} +  
                \Q{2}   \dc{1}   \L{1}   \Nc{2} +  
                \Q{2}   \dc{2}   \L{1}   \Nc{1} +  
                \L{1}   \L{2}   \Nc{1}   \ec{2}  ] \p{56} + \non\\  
&&<\p{12}\Hs{38}> [
                \Q{3}   \dc{2}  \L{1}   \Nc{2}  +         
                \L{1}   \L{2}   \ec{3}  \Nc{2}   ]\Hs{20} +  \non\\
&&<\p{23} \Hs{30}> [
                \h{1}  \Q{3}   \dc{3}   \pb{23} +          
                \h{1}  \L{3}   \ec{3}   \pb{23} +
                \Q{1}   \Q{2}   \uc{1}   \dc{2} +            
                \Q{1}   \uc{1}   \L{2}   \ec{2} +         
                \Q{1}   \uc{2}   \L{2}   \ec{1} +           
                \Q{2}   \dc{1}   \L{1}   \Nc{2} + \non\\           
&&\phantom{<\p{23}\Hs{30}> [}
                \Q{2}   \dc{2}  \L{1}   \Nc{1}  +           
                \L{1}   \L{2}   \ec{2}  \Nc{1}   ] \Hs{29}   + \non\\
&&<\p{4}  \Hs{30}> [
                \Q{3}   \uc{1}  \L{3}   \ec{1}  +            
                \Q{3}   \dc{1}  \L{3}   \Nc{1}  +             
               \uc{1}  \uc{1}  \dc{3}  \ec{3}  +             
                \uc{1}  \uc{3}  \dc{3}  \ec{1}  + 
               \uc{3}  \dc{1}  \dc{1}  \Nc{3}  +             
                \uc{3}  \dc{1}  \dc{3}  \Nc{1}   ] \Hs{29} + \non\\
&&<\p{4} \Hs{38}> [  
                  \hb{1}  \L{3}   \Nc{1} \ppb{56}  \Hs{20}   +  
                  \L{2}   \L{3}   \ec{1} \Hs{29}   \Vs{32} ] + \non\\ 
&&<\pp{4} \Hs{15}> [\Q{2}  \dc{1}  \L{2} + 
                    \uc{2} \dc{1}  \dc{2}  ] \Hs{19} \Vs{32}  + \non\\
&&<\pp{4} \Hs{38}> \uc{1}   \dc{2}   \dc{3} \Hs{29}   \Vs{32}   + \non\\
&&<\pb{4} \Hs{30}> [
                     \Q{1}   \Q{1}   \Q{3}    \L{3}  +           
                     \Q{1}   \Q{1}   \dc{3}   \uc{3} +            
                     \L{1}   \L{1}   \ec{3}   \Nc{3}  ] \Hs{29} + \non\\
&&<\pb{56} \Hs{30}> \h{1} [ \Q{3}   \dc{3} +          
                            \L{3}   \ec{3}   ] \p{56} \Hs{29} + \non\\
&&<\Hs{15} \Hs{30}> [
                      \Q{1}   \Q{2}   \Q{3}   \L{3}  +                
                      \Q{1}   \Q{2}   \dc{3}  \uc{3} +                
                      \Q{1}   \Q{3}   \uc{2}  \dc{3} +                
                      \Q{1}   \uc{2}  \L{3}   \ec{3} +                
                      \Q{1}   \uc{3}  \L{3}   \ec{2}   ] \p{3} + \non\\               
&&<\H{15} \H{30}> [  \Q{1}   \dc{2}  \L{3}   \Nc{3} +                
                      \Q{1}  \dc{3}  \L{2}   \Nc{3} +                
                      \Q{1}  \dc{3}  \L{3}   \Nc{2}   ] \p{56} 
\label{effw5ssm}
\eeqn

\no $W_{6}$:
\beqn
&& \h{3}  \hb{1}   [ \Q{1}   \uc{3}  \L{3}   \ec{1} + \Q{3}   \dc{1}  \L{1}   \Nc{3} ]+ \non\\       
&& [\Q{1}   \dc{1}   \L{2} +  \L{1}   \L{2}   \ec{1} ] \Hs{19} \Hs{29} \Vs{32}   + \non\\
&&<\Hs{30}> \h{1}   \h{1}   \hb{1}  [ \Q{3}    \dc{3} + \L{3}  \ec{3}  ] \Hs{29}  +  \non\\         
&&\ch h^{'} \h{3} [          
                \Q{2}   \Q{3}    \dc{2}  \dc{3} + 
                \Q{2}   \dc{2}   \L{3}   \ec{3} +   
                \Q{2}   \dc{3}   \L{2}   \ec{3} + 
                \Q{3}   \dc{2}   \L{3}   \ec{2} +  \non\\
&&\phantom{\ch h^{'} \h{3} [}
                \Q{3}   \dc{3}   \L{2}   \ec{2} +  
                \uc{2}   \dc{2}  \dc{3}  \ec{3} + 
                \uc{3}  \dc{2}  \dc{3}   \ec{2} +  
                \L{2}   \L{3}    \ec{2}  \ec{3}   ] + \non\\ 
&&\ch h^{'}  \hb{1} [
                \Q{1}   \Q{2}   \uc{1}   \dc{2} + 
                \Q{1}   \uc{1}   \L{2}   \ec{2} + 
                \Q{1}   \uc{2}   \L{2}   \ec{1} + 
                \Q{2}   \dc{1}   \L{1}   \Nc{2} + \non\\
&&\phantom{\ch h^{'}  \hb{1} [}
                \Q{2}   \dc{2}   \L{1}   \Nc{1} + 
                \L{1}   \L{2}   \ec{2}   \Nc{1}   ] + \non\\ 
&&<\pb{4}> \h{3}   \hb{1} [   
                \Q{1}   \Q{1}   \Q{3}    \L{3}  + 
                \Q{1}   \Q{1}   \uc{3}   \dc{3} + 
                \L{1}   \L{1}   \ec{3}   \Nc{3}   ] + \non\\
&&<\p{4}> \h{3}   \hb{1}   [   
                \Q{3}    \uc{1}   \L{3}   \ec{1} + 
                \Q{3}    \dc{1}   \L{3}   \Nc{1} + 
                \uc{3}   \dc{1}   \dc{1}  \Nc{3} + 
                \uc{3}   \dc{1}   \dc{3}  \Nc{1} + 
                \uc{1}   \uc{1}   \dc{3}  \ec{3} +
                \uc{1}   \uc{3}   \dc{3}  \ec{1}   ] + \non\\
&&\ch h^{'}  \h{3} [\Q{2} \dc{2} +  \L{2}   \ec{2}]
  [ \pb{23}   \Vs{31}   \Vs{32}  +  \ppb{56}  \Hs{19}   \Hs{20} ] + \non\\  
&&<\pb{56}> [ \h{3} \Q{2}  \dc{2} + \h{3} \L{2}   \ec{2}] \p{56} \Vs{31} \Vs{32} + \non\\
&&<\Hs{30}> [ (\h{1} \Q{2}   \dc{2} + \h{1} \L{2}  \ec{2}) \Vs{31} \Vs{32} +
              (\h{1} \Q{3}   \dc{3} + \h{1} \L{3}  \ec{3}) (\p{2}\p{2}+\p{3}\p{3})
                   ]\Hs{29} + \non\\            
&& <\Hs{38}> [\hb{1} \L{1}   \Nc{3}   \p{2}   \ppb{56} \Hs{20} ] + \non\\         
&&<\p{12}> [ \Q{1}   \Q{2}   \uc{1}   \dc{2} + 
             \Q{1}   \uc{1}   \L{2}   \ec{2} +    
             \Q{1}   \uc{2}   \L{2}   \ec{1} + \non\\
&&\phantom{<\p{12}> [ } 
             \Q{2}   \dc{1}   \L{1}   \Nc{2} +  
             \Q{2}   \dc{2}   \L{1}   \Nc{1} +   
             \L{1}   \L{2}    \ec{2}  \Nc{1}   ] [\p{2}\p{2}+\p{3}\p{3}]  +  \non\\
&&[<\pp{4}>  \uc{1} \dc{1} \dc{2} +  <\ppb{4}> \Q{1}   \dc{2}   \L{1} ]  
   \Hs{19} \Hs{29} \Vs{32}  + \non\\
&&<\Hs{15}> [   (\Q{2}   \dc{2}   \L{1}   +  \L{1}   \L{2}   \ec{2} ) \p{2} +  
                (\Q{3}   \dc{3}   \L{1}   +  \L{1}   \L{3}   \ec{3} ) \pb{23} ]    
                \Hs{19}   \Vs{32}   + \non\\
&&<\Hs{38}> [   \Q{1}   \dc{2}   \L{3} +   
                \Q{1}   \dc{3}   \L{2}   ] \p{3}   \Hs{29}   \Vs{32}      
\label{effw6ssm}
\eeqn

\newpage
\no {\it Mixed MSSM--Hidden Terms}

\no $W_{5}$:
\beqn
&& 
<\p{12}> [  \h{3}   \Q{2}   \dc{1}   \V{10}   \V{19} +  
            \h{3}   \L{1}   \ec{2}   (\V{ 4}   \V{13} +  \V{ 5}   \V{15}) ]+ \non\\
&&<\Hs{31}> \h{1}   \hb{1}  \Hs{19}  \V{19}   \V{37} + \non\\  
&&<\ppb{4} \Hs{30}> \h{1}   \hb{1}   \Vs{31} \H{23} \V{39}   + 
  <\ppb{4} \Hs{31}> \h{1}   \hb{1}   \Hs{19} \V{15} \V{40}   +  \non\\       
&&<\p{12} \pp{4}  > \h{3}   \Q{2}   \dc{1}   [\V{4}   \V{13} +  \V{5}   \V{15} ] +  
  <\p{12} \ppb{4} > \h{3}   \L{1}   \ec{2}   \V{10}   \V{19} + \non\\
&&<\p{12}   \Hs{38}> [
                (\Q{1}   \dc{2}   \L{3} + \Q{1}   \dc{3}  \L{2}  ) \H{25}   \V{35} +   
                (\uc{1}  \dc{2}  \dc{3} + \L{2}   \L{3}  \ec{1}  ) \H{23}   \V{39} ] + \non\\
&&<\p{23}  \Hs{30}> \Nc{2} [ \H{26}   \H{26}   \H{25}   \V{19} +  
                             \H{26}   \V{15}   \H{25}   \H{28} +  
                             \H{23}   \H{26}   \H{28}   \V{19} +
                             \H{23}   \V{15}   \H{28}   \H{28}]+  \non\\
&&<\Hs{30} \Hs{30}> \Nc{2} \p{56}  \Vs{31}  \V{19}  \V{35} +    \non\\
&&<\Hs{30} \Hs{38}> [\hb{1}  \Q{1}   \uc{3}  \V{19}  \V{35}  +    
                   \hb{1}  \Q{3}   \uc{1}  \V{15}  \V{39} ]+  \non\\
&&<\Hs{31} \Hs{38}> [\hb{1}  \L{1}   \Nc{3}  \V{19}  \V{37}  +
                   \hb{1}  \L{3}   \Nc{1}  \V{15}  \V{40}  +  \non\\
&&<\Hs{38} \Hs{38}> \hb{1}  \L{2}  \Vs{32}  \V{15}   \V{39}     
\label{effw5smh}
\eeqn

\no $W_{6}$:
\beqn
&&[     \Q{3}   \dc{2}   \L{3}   + 
        \Q{3}   \dc{3}   \L{2}   + 
        \uc{3}  \dc{2}   \dc{3}  + 
        \L{2}   \L{3}    \ec{3}  ] \Hs{19} \H{25}   \V{37}      + \non\\
&&<\p{12}> [ (\Q{1}   \dc{1}  \L{2} + \L{1}   \L{2}   \ec{1})  \Hs{19} \H{25}  \V{35} +
              \uc{1}  \dc{1}  \dc{2}  \Hs{19} \H{23}  \V{39} + \non\\
&&\phantom{<\p{12}> [}       
            \h{3}   \Q{2}   \dc{1}  \p{3}   \V{10}  \V{19} +     
            \h{3}   \L{1}   \ec{2}  \p{3}   \V{5}   \V{15}  ] + \non\\    
&&<\p{23}> [\Q{2}   \dc{2}  \L{2} + \uc{2}  \dc{2}  \dc{2} + 
            \L{2}   \L{2}   \ec{2}  ] \Hs{19} \H{25}  \V{37} + \non\\       
&&[ <\pb{4}> (\Q{3}   \dc{2}   \L{3}  + \uc{3}  \dc{2}  \dc{3})+
    <\ppb{4}>(\Q{3}   \dc{3}   \L{2}  + \L{2}   \L{3}   \ec{3})]
     \Hs{19}   \H{23}   \V{40} + \non\\
&&<\Hs{15}>       \h{3}    \Q{1}     \dc{3}   \Vs{32}   \H{28} \V{10}   + \non\\
&&<\Hs{38}>  \hb{1}   \L{3}  [\V{4}    \V{13}   \H{23}   \V{40} +
                              \H{23}   \V{5}    \V{15}   \V{40} +
                              \H{25}   \V{10}   \V{19}   \V{37} + 
                              \Vs{1}   \Vs{12}  \H{25}   \V{37}]\non\\  
\label{effw6smh}
\eeqn

\newpage
\no {\it Hidden Terms}

\no $W_{3}$:
\beqn
&& [<\Hs{30}>+ < \p{4}\pb{4}  \Hs{30}> + < \pp4 \ppb{4} \Hs{30}> ] \Vs{31} \H{25} \V{35} + \non\\
&& [<\Hs{31}>+ < \p{4} \pb{4}  \Hs{31}> + < \pp4 \ppb{4} \Hs{31}> ] \Hs{19} \V{19} \V{37} + \non\\ 
&& [<\ppb{4} \Hs{31}> +  
    <\p{4}   \pb{4}   \ppb{4}  \Hs{31}> +   
    <\pp{4}  \pb{4}   \pb{4}   \Hs{31}> +  
    <\pp{4}  \ppb{4}  \ppb{4}  \Hs{31}> ] \Hs{19} \V{15} \V{40} + \non\\ 
&& [<\ppb{4} \Hs{30}> + 
    <\p{4}   \pb{4}   \ppb{4}  \Hs{30}> +   
    <\pp{4}  \pb{4}   \pb{4}   \Hs{30}> +  
    <\pp{4}  \ppb{4}  \ppb{4}  \Hs{30}> ] \Vs{31} \H{23} \V{39} \non\\   
\label{effw3h}
\eeqn

\no $W_{4}$:
\beqn
&&                \Hs{37} \Vs{32} \H{26} \V{40}  + \non\\    
          && 
    <\p{4}> \Hs{37}   \Vs{32}   \H{28}   \V{37}  +  
    <\Hs{30}> \p{3}   \Vs{31}   \H{25}   \V{35}  + \non\\ 
          && 
    <\p{23} \pb{56}> [ \H{23}\H{23}\H{28}\H{28} + 
                       \H{23}\H{26}\H{25}\H{28} + 
                       \H{26}\H{26}\H{25}\H{25}]+ \non\\ 
          &&
    [<\p{23} \Hs{31}> \pb{23} + <\pb{56}\Hs{31}> \p{56} ] \Hs{19}\V{19}\V{37} + \non\\ 
          &&
    <\ppb{4}\Hs{30}>  \p{3}\Vs{31}\H{23}\V{39} + \non\\
          &&
    <\Hs{15}\Hs{30}> \Hs{20}\Vs{31}\H{25}\V{10}+ \non\\ 
          && 
      [<\pb{4}   \p{4}   \Hs{30}> + <\ppb{4}   \pp{4}   \Hs{30}> ] \p{3} \Vs{31}   \H{25}   \V{35} +  \non\\
          && 
      [<\ppb{4}   \pb{56}   \Hs{31}> \p{56}    +  <\ppb{4}   \p{23}   \Hs{31}> \pb{23}  ]
         \Hs{19}   \V{15}   \V{40} +  \non\\
          && 
        <\pp{4}   \Hs{15}   \Hs{30}> \Hs{20}  \Vs{31}  \H{23}   \V{5} 
\label{effw4ahid}
\eeqn

\no $W_{5}$:
\beqn
&&    \p{2} \Hs{37}  \Vs{32}  \H{26}   \V{40}  + \non\\  
  && < \Hs{30}> (\p{2} \p{2}+ \p{3}\p{3})\Vs{31} \H{25} \V{35}  +   
     < \Hs{31}> (\p{2} \p{2}+ \p{3}\p{3})\Hs{19} \V{19} \V{37}  + \non\\ 
          && 
<\ppb{4}   \Hs{30}> (\p{2}   \p{2}  +  \p{3}  \p{3} )  \Vs{31}   \H{23}   \V{39} +
<\ppb{4}   \Hs{31}> (\p{2}   \p{2}  +  \p{3}  \p{3} )  \Hs{19}   \V{15}   \V{40} + \non\\
         &&
 <\Hs{15} \Hs{30}> \p{2}  \Hs{20}  \Vs{31}  \H{25}  \V{10} +     
 [<\p{23} \Hs{30}> \p{3}  \pb{23} + <\pb{56}\Hs{30}> \p{3}  \p{56} ] \Vs{31}   \H{25}  \V{35}    \non\\
\label{effw5hid}
\eeqn

\no $W_{6}$:
\beqn
&& <\pb{56}>    \Vs{31}   \Vs{32} [ \H{26}   \H{26}   \H{25}   \H{25} +   
                                    \H{23}   \H{26}   \H{25}   \H{28} + 
                                    \H{23}   \H{23}   \H{28}   \H{28}]+ \non\\
         &&
<\Hs{30}> (\p{2}     \p{2}   \p{3} +  \p{3}     \p{3}   \p{3}) \Vs{31} \H{25}   \V{35}         
\label{effw6}
\eeqn

\newpage
\vskip 0.5truecm
\no\underline{Flat direction 1V modifications to superpotential, $W^{\rm FD1V}$:}

\no {\it Singlet Terms}

\no $W_{3}$:
\beqn
[<\pb{12}   \Hs{30}>  + < \p{12}   \pb{12}  \pb{12}   \Hs{30}>  +
   <\pb{12}   \p{4}    \pb{4}    \Hs{30}>  +  
   <\pb{12}   \pp{4}   \ppb{4}   \Hs{30}>] \Hs{29}   \Vs{31}   \Vs{32}\non\\  
\label{m1effw73s}
\eeqn

\no $W_{4}$:
\beqn
&&
<\pb{12}    \p{23}  \Hs{30}> [\pb{23}  \Hs{29}  \Vs{31} \Vs{32} +
                              \ppb{56} \Hs{19}  \Hs{20} \Hs{29}]+ 
<\pb{12}   \pb{56}  \Hs{30}>  \p{56}   \Hs{29}  \Vs{31} \Vs{32} 
\label{m1effw4f7sig}
\eeqn

\no $W_{5}$:
\beqn
<\pb{12}   \Hs{30}>[   \p{2}   \p{2}  + \p{3}   \p{3}   ]  \Hs{29} \Vs{31} \Vs{32}       
\label{m1effw5f7sig}
\eeqn

\no {\it MSSM Terms}

\no $W_{3}$:
\beqn
&&[ <\Hs{30}> + <\p{4} \pb{4} \Hs{30}> ] \hb{4}   \Q{3}   \uc{3} + \non\\    
&&[<\p{12} \Hs{31}> +  <\p{12} \p{12}\pb{12}\Hs{31}> +
   <\p{12} \p{4}  \pb{4} \Hs{31}> + <\p{12} \pp{4}  \ppb{4} \Hs{31}>]
 [ \hb{4} \Q{1}   \uc{1} + \hb{4} \L{1} \Nc{1} ]+ \non\\    
&&<\p{23} \Hs{15}\Hs{30}>  \chb \bh^{1} \Q{1}  \uc{2} + 
 <\p{4}  \Hs{15} \Hs{38}>  \chb \bh^{1} \L{3}  \Vs{32}
\label{m1effw3fmssm}
\eeqn

\no $W_{4}$:
\beqn
&&[ <\Hs{31}> + <\p{12} \pb{12} \Hs{31}> + <\p{4}  \pb{4}  \Hs{31}>+ 
    <\pp{4} \ppb{4} \Hs{31}]  \h{3}  \hb{4} \Vs{31} \Vs{32} + \non\\
&&<\p{23} \Hs{30}> \hb{4}\Q{3}\uc{3}\pb{23} + 
  <\pb{56}\Hs{30}> \hb{4}\Q{3}\uc{3}\p{56}  +
  <\Hs{15}\Hs{31}> \hb{4}\L{1}\Hs{19}\Vs{32} + \non\\
&&<\p{12}    \p{12}   \pb{12}>[ 
                              \Q{1}   \Q{2}   \uc{1}   \dc{2}   + 
                              \Q{1}   \uc{1}   \L{2}   \ec{2}   +
                              \Q{1}   \uc{2}   \L{2}   \ec{1}   + 
                              \Q{2}   \dc{1}   \L{1}   \Nc{2}   + \non\\
&&\phantom{<\p{12}   \p{12}   \p{12}>[} 
                              \Q{2}   \dc{2}   \L{1}    \Nc{1}   +
                              \L{1}   \L{2}    \ec{2}   \Nc{1} ] + \non\\
&&<\pb{12} \Hs{30}>\ch  h^{'}  [\Q{3} \dc{3} + \L{3} \ec{3} ] \Hs{29}+\non\\
&&<\p{23}  \Hs{30}>\chb \bhp[\Q{1} \uc{1} + \L{1} \Nc{1} ] \Hs{29}+\non\\
&&<\p{12}   \p{23}\Hs{31}>[\hb{4}   \Q{1}   \uc{1} +  
                           \hb{4}   \L{1}   \Nc{1}  ] \pb{23} + \non\\
&&<\p{12}  \pb{56} \Hs{31}>[\hb{4}   \Q{1}   \uc{1} +  
                           \hb{4}   \L{1}   \Nc{1}  ] \p{56} 
\label{m1effw4f7ssm}
\eeqn

\no $W_{5}$:
\beqn
&&<\Hs{30} > [ \h{1} \hb{1}\hb{4}\Q{3}\uc{3} + 
               \hb{4}\Q{3}\uc{3}(\p{2}\p{2}+\p{3}\p{3})+
               \hb{4}\L{3}\Nc{3}\p{56}\ppb{56} ] + \non\\
&&  <\pb{56} \Hs{31} > \h{3} \hb{4}  \p{56} \Vs{31}\Vs{32} +       
    <\Hs{30} \Hs{31} > \h{1} \hb{4}  \Hs{29}\Vs{31}\Vs{32} + \non\\      
&&<\p{12}   \pb{12}>[\h{3} \Q{2} \dc{2} + \h{3} \L{2} \ec{2}  ] \Vs{31}   \Vs{32} + \non\\
&&  <\p{12} \Hs{31}>[ \h{1}  \hb{4} \ppb{56} \Hs{19} \Hs{20}  +     
                     (\hb{4} \Q{1}  \uc{1} + \hb{4}  \L{1}  \Nc{1} ) 
                     (\p{2} \p{2} + \p{3} \p{3} )] + \non\\
&&  <\Hs{31}>[ \ch h^{'}  \hb{4}  \pb{23} \Vs{31} \Vs{32}+     
               \ch h^{'}  \hb{1}  \hb{4}  \Q{1}   \uc{1} +    
               \ch h^{'}  \hb{1}  \hb{4}  \L{1}   \Nc{1} +    
               \ch h^{'}  \h{3}   \hb{4}  \Q{3}   \dc{3} +    
               \ch h^{'}  \h{3}   \hb{4}  \L{3}   \ec{3}  ] \non\\    
\label{m1effw5f7ssm}
\eeqn

\no $W_{6}$:
\beqn
&& <\Hs{30}>[\hb{4}  \Q{2}    \uc{2}  \ppb{56} +     
             \hb{4}   \L{2}   \Nc{2}  \p{56}     ]\Vs{31}\Vs{32} \non\\   
&& <\Hs{31}>[ \h{1} \h{3}  \hb{1}  \hb{4}  +     
              \h{3} \hb{4} (\p{2}   \p{2}  +  \p{3} \p{3} ) ]\Vs{31}\Vs{32} +
\non\\
&&   <\Hs{38}> \hb{1}  \hb{4} \Q{1}   \uc{3}  \L{2}  \Vs{32}  
\label{m1effw6f7ssm}
\eeqn

\no {\it Mixed MSSM--Hidden Terms}

\no $W_{4}$:
\beqn
&&[<\p{12} > +  <\p{12}  \p{12} \pb{12}> + <\p{12}  \p{4}  \pb{4} > +       
  <\p{12}  \pp{4} \ppb{4}> ] \hb{4}   \L{1}   \H{23} \V{5} + \non\\
&& <\p{12}\ppb{4}>  \hb{4}   \L{1}   \H{25} \V{10}
\label{m1effw4f5mix}
\eeqn

\no $W_{5}$:
\beqn
&&[<\pb{12} \p{23}> \hb{1} +   
   < \p{23}\Hs{31}> \hb{4} ] \L{2} \Hs{19} \H{25}  \V{37} + \non\\
&&[<\p{12}  \p{23}> \pb{23} +       
   <\p{12} \pb{56}> \p{56}  +       
   <\p{23} \Hs{30}> \Hs{29}  ]\hb{4}  \L{1}   \H{23}  \V{5}       
\label{m1effw5f7mix}
\eeqn

\no $W_{6}$:
\beqn
 [<\p{12}>\h{1}  \hb{1} + <\p{23}>  \h{3}  \hb{1} + 
  <\p{12}> (\p{2}\p{2} + \p{3} \p{3} )] 
  \hb{4}  \L{1}   \H{23}  \V{5}     
\label{m1effw6f7mix}
\eeqn

\no {\it Hidden Terms}

\no $W_{3}$:
\beqn
&&<\p{12}   \pb{12}   \Hs{30}>  \Vs{31}   \H{25}   \V{35} +  
  <\p{12}   \pb{12}   \Hs{31}>  \Hs{19}   \V{19}   \V{37} +   \non\\   
&&<\p{12}   \pb{12}   \ppb{4}   \Hs{30}>  \Vs{31}   \H{23}   \V{39} +    
  <\p{12}   \pb{12}   \ppb{4}   \Hs{31}>  \Hs{19}   \V{15}   \V{40}    
\label{m1effw53h}
\eeqn

\no $W_{4}$:
\beqn
<\p{12}   \pb{12}   \Hs{30}> \p{3}   \Vs{31}   \H{25}   \V{35}
\label{m1effw4h7}
\eeqn

\newpage
\no\underline{Flat direction 2$^{'}$ modifications to superpotential, $W_{{\rm FD2}^{'}}$:}

\no {\it MSSM Terms}

\no $W_{3}$:
\beqn
  <\Hs{15} \Hs{31} \pp{56}>  h^{'} \Q{1}  \dc{2}          
\label{p2effw3mssm7}
\eeqn

\no $W_{4}$:
\beqn
&& <\p{12} \pp{56} \ppb{56}>[\Q{1}   \Q{2}   \uc{1}  \dc{2}+     
                              \Q{1}   \uc{1}  \L{2}   \ec{2}+     
                              \Q{1}   \uc{2}  \L{2}   \ec{1}+     
                              \Q{2}   \dc{1}  \L{1}   \Nc{2}+ \non\\     
&&\phantom{<\p{12} \pp{56} \ppb{56}>[}
                              \Q{2}   \dc{2}  \L{1}   \Nc{1}+     
                              \L{1}   \L{2}   \ec{2}  \Nc{1} ]\non\\     
&&<\pp{56} \Hs{15} \Hs{30}> [
                          \Q{1}   \Q{3}   \Q{3}   \L{2}  +        
                           \Q{1}   \Q{3}   \uc{3}  \dc{2} +      
                           \Q{1}   \uc{3}  \L{2}   \ec{3} ] + \non\\      
&&<\pp{56} \ppb{56}\Hs{30}> [ \h{1}   \Q{3}   \dc{3} +    
                              \h{1}   \L{3}   \ec{3}] \Hs{29}      
\label{p2effw4mssm7}
\eeqn

\no $W_{5}$:
\beqn
  <\ppb{56} \pp{56}>[ \h{3}  \Q{2}   \dc{2} +       
                      \h{3}  \L{2}   \ec{2}   ] \Vs{31}\Vs{32}      
\label{p2effw5mssm7}
\eeqn

\no {\it Hidden Terms}

\no $W_{3}$:
\beqn
&&   <\ppb{4} \pp{56} \ppb{56} \Hs{30}>\Vs{31}  \H{23}  \V{39} +    
     <\ppb{4} \pp{56} \ppb{56} \Hs{31}>\Hs{19}  \V{15}  \V{40}   
\label{p2effw4h7}
\eeqn

\vskip 0.5truecm
\no\underline{Flat direction 2V modifications to superpotential, $W^{\rm FD2V}$:} 


\no {\it MSSM Terms}

\no $W_{4}$:
\beqn
<\Hs{30} \ppb{56}> \hb{4}\L{3}\Nc{3} \p{56}
\label{m2effw4ssm7}
\eeqn

\no $W_{5}$:
\beqn
<\Hs{30} \ppb{56}> \hb{4}\Q{2}\uc{2} \Vs{31}\Vs{32}  
\label{m2effw5ssm7}
\eeqn



\vskip 0.5truecm
\no\underline{Flat Direction $2^{'}V$ modifications to superpotential:} 

\no {\it Singlet Terms}

\no $W_{3}$:
\beqn
&&<\pb{12}  \pp{56}   \ppb{56}  \Hs{30}> \Hs{29}   \Vs{31}   \Vs{32}         
\label{m2effw3f7sig}
\eeqn


\no {\it MSSM Terms}

\no $W_{3}$:
\beqn
&& <\Hs{30}\pp{56} \ppb{56}> \hb{4}\Q{3}\uc{3} + \non\\
&& <\p{12} \pp{56} \ppb{56} \Hs{31} > [ \hb{4} \Q{1} \uc{1} + \hb{4} \L{1} \Nc{1} ]      
\label{m2effw3f7ssm}
\eeqn

\no $W_{4}$:
\beqn
<\Hs{30} \pp{56} \ppb{56}> \hb{4}\Q{3}\uc{3} 
\label{m2effw4f7ssm}
\eeqn

\no {\it Mixed MSSM--Hidden Terms}

\no $W_{5}$:
\beqn
 <\p{12}\ppb{56}>\hb{4}  \L{1}   \H{23}  \V{5} \pp{56}    
\label{m2effw5smh7}
\eeqn

\vskip 0.5truecm
\no\underline{Flat direction 3 modifications to superpotential, $W^{\rm FD3}$:}

\no {\it MSSM Terms}

\no $W_{4}$:
\beqn
&&<\Hs{15}   \Hs{19}>   [ \Q{2}   \dc{2}  \L{1}  +
                          \L{1}   \L{2}   \ec{2}   ] \Vs{32} + \non\\ &&
<\pp{4} \Hs{15} \Hs{19}>[ \Q{2}   \dc{1}  \L{2}  +        
                           \uc{2}  \dc{1}  \dc{2}   ] \Vs{32}       
\label{g3meffw4f6ssm}
\eeqn

\no $W_{5}$:
\beqn
&&<\Hs{19}> [ \Q{1}   \dc{1} \L{2} +  
              \L{1}   \L{2}  \ec{1} ] \Hs{29}  \Vs{32} +\non\\ 
     &&
   <\pp{4} \Hs{19}> \uc{1}  \dc{1}  \dc{2} \Hs{29} \Vs{32} +     
   <\ppb{4}\Hs{19}> \Q{1}   \dc{2}  \L{1}  \Hs{29} \Vs{32} + \non\\     
     &&
   <\Hs{15}\Hs{19}>[\Q{2}   \dc{2}  \L{1}   \p{2} +      
                    \Q{3}   \dc{3}  \L{1}  \pb{23}+      
                    \L{1}   \L{2}   \ec{2}  \p{2} +      
                    \L{1}   \L{3}   \ec{3} \pb{23} ] \Vs{32} \non\\      
\label{g3meffw5f6ssm}
\eeqn

\no $W_{6}$:
\beqn
   <\Hs{19}>[\Q{1}   \dc{1}  \L{2} +      
             \L{1}   \L{2}  \ec{1}  ] (\p{2}+\p{3})\Hs{29} \Vs{32}      
\label{g3meffw6f7ssm}
\eeqn

\no {\it Mixed MSSM--Hidden Terms}
\no $W_{5}$:
\beqn
   <\p{12} \Hs{19}>[ \Q{1}   \dc{1}  \L{2} + 
                     \uc{1}  \dc{1}  \dc{2}+ 
                     \L{1}   \L{2}   \ec{1} ]  \H{25}  \V{35} 
\label{g3meffw5f6smh}
\eeqn





\vskip 0.5truecm
\no\underline{Flat Direction $3V$ modifications to superpotential, $W^{\rm FD3V}$:} 

\no {\it MSSM Terms}

\no $W_{3}$:

\beqn
   <\Hs{15} \Hs{19} \Hs{31}> \hb{4} \L{1} \Vs{32}      
\label{v3effw3f6ssm}
\eeqn    

\no $W_{4}$:

\beqn
   <\Hs{15} \Hs{19} \Hs{31}> \hb{4} \L{1} \Vs{32} \p{2}    
\label{v3effw4f7ssm}
\eeqn

\vskip 0.5truecm
\no\underline{Flat Direction $4$ modifications to superpotential $W^{\rm FD4}$:} 

\no {\it MSSM Terms}

\no $W_{4}$:
\beqn
<\p{12}  \Hs{38}  \Hs{20}> [\Q{3}   \dc{2}   \L{1}   \Nc{2}+                 
                                   \L{1}   \L{2}    \ec{3}  \Nc{2} ]               
\label{m4effw4f7ssm}
\eeqn    






               
\vfill\eject

\bigskip
\medskip

\def\bibiteml#1#2{ }
\bibliographystyle{unsrt}

\hfill\vfill\eject
\end{document}